\documentclass{article}

  \PassOptionsToPackage{numbers, compress}{natbib}

     \usepackage[final]{neurips_2021}

\usepackage[utf8]{inputenc} 
\usepackage[T1]{fontenc}    
\usepackage{hyperref}       
\usepackage{url}            
\usepackage{booktabs}       
\usepackage{amsfonts}       
\usepackage{nicefrac}       
\usepackage{microtype}      
\usepackage{xcolor}         

\usepackage{amsmath,amssymb,amsthm,bm,mathtools,graphicx,subfig,color,tcolorbox,algorithm,algorithmic}
\usepackage{caption,fancyhdr,titlesec,indentfirst,booktabs,verbatim,hyperref,cases,titletoc,multirow,wasysym,empheq,setspace,natbib,chngcntr}

\usepackage{hyperref}
\hypersetup{colorlinks,linkcolor={blue},citecolor={blue},urlcolor={red}} 
\setcitestyle{square, numbers}

\newtheorem{theorem}[]{Theorem}

\newtheorem{proposition}[]{Proposition}

\def\E{\mathbb E}

\def\({\left(}
\def\){\right)}
\def\[{\left[}
\def\]{\right]}

\usepackage{xr}
\makeatletter
\newcommand*{\addFileDependency}[1]{
  \typeout{(#1)}
  \@addtofilelist{#1}
  \IfFileExists{#1}{}{\typeout{No file #1.}}
}
\makeatother

\title{Learning in Multi-Stage Decentralized Matching Markets}

\author{%
  Xiaowu Dai \\
  UC Berkeley\\
  \texttt{xwdai@berkeley.edu} \\
  \And
  Michael I. Jordan \\
  UC Berkeley\\
  \texttt{jordan@cs.berkeley.edu} \\
}

\begin{document}

\maketitle

\begin{abstract}
Matching markets are often organized in a multi-stage and decentralized manner.  Moreover, participants in real-world matching markets often have uncertain preferences. This article develops a framework for learning optimal strategies in such settings, based on a nonparametric statistical approach and variational analysis.  We propose an efficient algorithm, built upon concepts of ``lower uncertainty bound'' and ``calibrated decentralized matching,'' for maximizing the participants' expected payoff. 
We show that there exists a  welfare-versus-fairness trade-off that is characterized by the uncertainty level of acceptance.  Participants will strategically act in favor of a low uncertainty level to reduce competition and increase expected payoff.
We prove that participants can be better off with multi-stage matching compared to single-stage matching. 
We demonstrate aspects of the theoretical predictions through simulations and an experiment using real data from college admissions.
\end{abstract}

\section{Introduction}

Two-sided matching markets have played an important role in microeconomics for several decades \cite{rothsatomayor1990}.
Matching markets are used to allocate indivisible “goods” to
multiple decision-making agents based on mutual compatibility as assessed via sets of preferences. 
Such a market does not clear through prices. For example, a student applicant cannot simply demand the college she prefers but must also be chosen by the college. 
Matching markets are often organized in a decentralized way. Each agent makes their decision independently of others’ decisions, and each agent can have multiple stages of interactions with the other side of the market. College admissions with waiting lists and academic job markets are notable examples.
We refer to such markets as \emph{multi-stage decentralized matching markets}.

Uncertain preference is ubiquitous in multi-stage decentralized matching markets. For instance, colleges competing for students lack information on  students' preferences.
An admitted student may receive offers from other colleges. She needs to accept one or reject all offers within a short period during each stage of early, regular, and waiting-list admissions \cite{avery2003}.
This admission process provides little opportunity for colleges to learn students’ preferences, which are uncertain due to competition among colleges and variability in the relative popularity of colleges over time.  Such uncertain preferences pose a challenge for colleges in their attempt to formulate an  optimal admission strategy. 
Consequently, colleges  may end up enrolling too many or too few students relative to their capacity or having enrolled students overly far from the attainable optimum in quality.

This paper addresses the following two research questions:
(i) Given the uncertain preferences on one side of the market (e.g., students), how can agents (e.g., colleges) learn an optimal strategy that maximizes expected payoffs based on historical data?
(ii) What are the fundamental implications of multi-stage decentralized matching on the welfare and fairness for both sides of the market?
We study these two questions using \emph{nonparametric statistical methodology} and \emph{variational  analysis}. We propose a new algorithm for maximizing agents' expected payoffs that is based on learning stage-wise optimal strategies and calibrating state parameters based on historical data. In particular, our algorithm balances the opportunity cost and the penalty for exceeding the quota for calibration. 
Based on the calibrated state, the algorithm efficiently learns an optimal strategy using statistical machine learning methods.
The statistical model not only provides a foundation for the algorithm but it also provides an analytical framework for understanding the implications of the approach for welfare and fairness. 
We show that agents will favor arms with realistic and stable opportunities for matching instead of only targeting the top-ranked arms. 
Moreover, we show that agents are better off with multi-stage decentralized matching as compared to single-stage decentralized matching.

Adopting literature from the bandit literature, our model has a set of \emph{agents}, each with limited capacity, and a set of \emph{arms}. Each agent values two attributes of an arm: a ``score" that is common to all agents and a ``fit" that is agent-specific and independent across agents. Agents rank arms according to their scores and fits.  An agent's strategy consists of how many and which arms to pull at each stage.  On the other hand, there is no restriction on the preferences of arms.  The model allows uncertainty in the preferences, which is incorporated into the arms' stage-wise acceptance probabilities. The acceptance probability depends on the unknown state of the world and the competition of agents at each stage.  We consider a simple timeline for multi-stage markets. At each stage, agents simultaneously pull sets of arms. Each arm accepts at most one of the agents that pulled it. The arms have to make irreversible decisions at each stage without knowing which other agents might select them in later stages.

\paragraph{Our contributions}
There are two main contributions in this paper, which correspond to the two questions above. Our first contribution is to propose a new algorithm that maximizes the agent’s expected payoff in multi-stage decentralized matching markets. 
The algorithm sequentially learns the optimal strategy at each stage and is built upon notions of \emph{lower uncertainty bound} (LUB) and \emph{calibrated decentralized matching} (CDM). 
The key idea is to calibrate the state parameter in a  data-driven approach and take the opportunity cost and penalty for exceeding the quota into account. The calibration can be performed under both average-case and worst-case metrics, depending on whether we are maximizing the averaged or minimal expected payoff with respect to the uncertain state. 
Given the calibrated state, the algorithm efficiently learns the optimal strategy using historical data via statistical machine learning methods.

The second contribution is providing an analytical framework for understanding the welfare and fairness implications. 
We show that agents favor arms with low uncertainty in levels of acceptance, suggesting that agents prefer arms with a realistic and stable chance for matching instead of only targeting the top-ranked arms. 
Such strategic behavior improves the agent's expected payoff since otherwise, by the time that arms have rejected that agent, the next-best arms that the agent has in mind may already have accepted other agents. 
However, the strategic behavior leads to unfair outcomes for arms because some arms are not pulled by their favorite agents even though these agents pull arms ranked below them.
We prove that agents are better off in multi-stage decentralized matching markets compared to single-stage decentralized matching markets. 

\paragraph{Related work}
This paper is related to three strands of literature. 
The first line is on matching markets. Most theoretical work on matching markets traces back to \cite{gale1962} that formulated a model of two-sided matching without side payments, and  \cite{shapley1971assignment} that formulated a model of two-sided matching with side payments.
The model in \cite{shapley1971assignment} is also related to the maximum weighted bipartite matching and its to stochastic and online generalizations \cite{mehta2013online}.
Our goal is to design algorithms for maximizing the agent's welfare under the model of \cite{gale1962}, given the  uncertain preferences of arms. This is different from the goal of finding a matching with the largest size in maximum matching literature \cite{shapley1971assignment, mehta2013online}. 
The second strand of literature is on the decentralized interactions in  matching markets  \cite{dai2020learning, das2005two,  diamantoudi2015decentralized, neiderle2009,  roth1997} and search literature
\cite{montgomery1991equilibrium, peters1991ex}.
Our paper contributes to this strand of literature via its analysis of multi-stage markets that allow uncertain preferences. We also study the economic implications for strategic behaviors in multi-stage decentralized markets.
The third related body of literature is on algorithmic studies of college admissions.
The celebrated work in \cite{gale1962} introduced the deferred acceptance algorithm implemented under central clearinghouses. 
Recent works have been focused on equilibrium admissions, 
students’ efforts, and students' information acquisition costs in forming preferences; see, \cite{azevedo2016, chade2014,chade2006,che2016,epple2006,fu2014,hafalir2018, immorlica2020information}. In contrast, we emphasize students' multidimensional abilities and multiple colleges  competing for students.
The students' preferences are uncertain due to the competition among colleges and variability in the relative popularity of colleges over time.
We develop a statistical model for learning the optimal strategies using historical data.


\section{Problem Formulation}
\label{sec:model}


\paragraph{Multi-stage decentralized  matching markets}
Let $\mathcal P=\{P_1,P_2,\ldots,P_m\}$ be a set of $m$ agents. Let  $\mathcal A=\{A_1, A_2, \ldots,A_n\}$ the a set of $n$ arms. Here $\mathcal P$ and $\mathcal A$ are the sets of participants on the two sides of the matching market. 
Each agent $P_i$ has a quota $q_i\geq 1$. We assume that $q_1+q_2+\cdots+q_m\leq n$. There are total of $K\geq1$ stages of the matching process. 
At each stage, an agent who has not used up its quota can pull available arms in the market. When multiple agents select the same arm, only one agent can successfully pull the arm according to the arm's preference.  We denote $[m]\equiv\{1,\ldots,m\},[n]\equiv\{1,\ldots,n\},$ and $[K]\equiv\{1,\ldots,K\}$.
Decentralized matching markets require participants to make their decisions independently of others’ decisions \cite{roth2008, roth1997}.
Notable examples of such markets include
college admissions  in the United States, Korea, and Japan, where  $\mathcal P$ and $\mathcal A$ represent the sets of colleges and students,  respectively \cite{avery2003, avery2010}.
Our goal is to learn the agent's optimal strategy  for maximizing the expected payoff. A strategy consists of deciding how many and which  arms to pull at each stage. 
Agent's decision-making in decentralized markets faces incomplete information about other agents' decisions and arms' preferences. 


\paragraph{Participants' preferences}
The agents' preferences are based on the arms' latent utilities. Consider the following latent utility model:
\begin{equation}
\label{eqn:defofutility}
U_i(A_j) = v_j + e_{ij},\quad\forall i\in [m], j\in[n],
\end{equation}
where $v_j\in[0,1]$ is arm $A_j$'s \emph{systematic score} considered by all agents, and $e_{ij}\in[0,1]$ is an agent-specific \emph{idiosyncratic fit} considered only by agent $P_i$, $i\in[m]$. 
A utility model with a similar separable structure has been widely used in the matching market literature \cite{ashlagi2019, che2016, dai2020learning}. 

The arms' preferences have no restrictions  and can involve uncertainty.
From an agent's perspective, arms accept offers with probabilities dependent on opponents' strategies and arms' preferences.
Let the parameter $s_{i,k}\in[0,1]$ be the \emph{state of the world} \cite{savage1972}  for agent $P_i$, such that the probability that an arm $A_j$ accepts $P_i$ at stage $k$ is
$\pi_{i,k}(s_{i,k},v_j),\forall i\in[m],j\in[n], k\in[K]$.
Since agents compete for arms with a higher score, the acceptance probability $\pi_{i,k}(s_{i,k}, v_j)$ models the agents' competition through the dependence on the score $v_j$.
Moreover, $\pi_{i,k}(s_{i,k}, v_j)$ incorporates the arm's uncertain preference into the state $s_{i,k}$.
It is known that there exists a valid probability mass function $\pi_{i,k}(s_{i,k},v_j)$ \cite{dai2020learning}. 
We assume that $\pi_{i,k}(s_{i,k}, v_j)$ is strictly increasing and continuous in $s_{i,k}$.
Thus, a larger  value of the state $s_{i,k}$ corresponds to the case that agent $P_i$ is more popular. 
In practice, the true state is  unknown a priori to $P_i$ and needs to be estimated from data. 
For instance,  the \emph{yield} in  college admissions is defined as the rate at which  a college's admitted students accept the offers. However, the yield is unknown a priori to the college in the current year \cite{che2016}. Colleges can only estimate the distribution of the yield from historical data. 
In this paper, we study a nonparametric model of $\pi_{i,k}(\cdot, \cdot)$ by assuming it belongs to a reproducing kernel Hilbert space (RKHS) \cite{aronszajn1950theory, wahba1990spline}. 
Later, in Section \ref{sec:LUBcdm}, we propose an algorithm for calibrating $s_{i,k}$ and efficiently estimating $\pi_{i,k}(\cdot, \cdot)$ using historical data.
Given the latent utility $U_i(A_j)$ and the acceptance probability $\pi_{i,k}(s_{i,k}, v_j)$, agent $P_i$'s expected utility  of pulling arm $A_j$  at stage $k$ is
$(v_j+e_{ij})\pi_{i,k}(s_{i,k}, v_j)$.

\paragraph{Timeline of the matching}
First, Nature draws a state such that arms' preferences are realized. Denote by $s_{i,k}^*$ the true state for agent $P_i$ at stage $k$.
Next, arms display their interests to all agents. 
For example, students apply to colleges in a given period. Under the assumption that students incur negligible application costs,  submitting applications to all colleges is the dominant strategy as students  lack information on how colleges evaluate their academic ability or personal essays \cite{avery2010, che2016}. 
Next, at each stage $k\in[K]$, agents simultaneously pull available arms that have not previously rejected them. Each arm either accepts one of the agents  that pulled it (if any) or rejects all. An arm exits the market once it accepts an agent, and agents are allowed to exit the market at any time.
The  arms act simultaneously at each stage. They
cannot ``hold" offers for accepting or rejecting at a later stage. Hence, agents make ``exploding" offers, and arms have to make \emph{irreversible} decisions without knowing what other offers are coming in later stages. 
Finally, this multi-stage matching process ends when all agents have exited or when a pre-specified number of stages has been reached. If there remain arms in the market when the matching has terminated,  these arms are unmatched. 


\paragraph{Agent's expected payoff}
An agent's goal is to maximize the expected payoff, which consists of two parts: the expected utilities and the penalty for exceeding the quota. 
Let $\mathcal A_{k}$ be the set of arms that are available in the market at stage $k\in[K]$.
Suppose that agent $P_i$ pulls arms from the set  $\mathcal B_{i,k}\subseteq \{\mathcal A_k\setminus \cup_{l\leq k-1}\mathcal B_{i,l}\}$ at stage $k$, where $\mathcal A\setminus \mathcal B$ denotes that set $\mathcal A$ minus set $\mathcal B$.
Let $\mathcal C_{i,k}\subseteq\mathcal B_{i,k}$ be the set of arms that accept $P_i$ at stage $k$.  Then $\mathcal C_{i,k}$ is unknown until  stage $k+1$, where $k\leq K-1$, and $\mathcal C_{i,K}$ is unknown until the end of the matching process.
Then $P_i$'s  expected payoff at stage $k\in[K]$ is lower bounded by
\begin{equation}
\label{eqn:totalutility}
\begin{aligned}
 \mathcal U_{i,k}[\mathcal B_{i,k}] & \equiv \sum_{j\in\mathcal B_{i,k}}\left(v_j+e_{ij}\right)\pi_{i,k}(s^*_{i,k},v_j) - \gamma_i\max\{\mathcal N_{i,k}(\mathcal B_{i,k})-q_i,\ 0\}.
\end{aligned}
\end{equation}
Here $\mathcal N_{i,k}(\mathcal B_{i,k}) \equiv \sum_{j\in\mathcal B_{i,k}}\pi_{i,k}(s^*_{i,k},v_j)+\text{card}(\cup_{l\leq k-1}\mathcal C_{i,l})$, and $s_{i,k}^*$ is the true state for agent $P_i$ at stage $k$. We assume that the marginal penalty $\gamma_i$ satisfies $\gamma_i>\max_{j\in\mathcal A}\{v_j+e_{ij}\}$, which implies that the penalty is greater than arm's latent utility.
Since our model involves unknown strategies of the opponents and uncertain arms' preferences, 
we consider the optimal expected payoff in (\ref{eqn:totalutility}) instead of the optimal realized payoff. Similar expected payoff have been studied in \cite{che2016,dai2020learning}.

\section{Statistical Learning of the Optimal Strategy}
\label{sec:multistage}

We consider a variational formulation of the optimal strategy in Section \ref{sec:modifiedutility} and propose a two-step algorithm using a statistical machine learning method in Section \ref{sec:LUBcdm}.
  
\subsection{Variational formulation}
\label{sec:modifiedutility}
The problem of finding the optimal set of arms, and the corresponding optimal value $\bar{\mathcal U}_{i}$, can be described as follows:
\begin{equation}
\label{eqn:optimalsolution}
    \bar{\mathcal U}_{i} = \underset{\mathcal B_{i,k}\subseteq \{\mathcal A_k\setminus \cup_{l\leq k-1}\mathcal B_{i,l}\},k\in[K]}{\max}\sum_{k\in[K]}\mathcal U_{i,k}[\mathcal B_{i,k}],
\end{equation}
where the expected payoff $\mathcal U_{i,k}$ is defined in (\ref{eqn:totalutility}).
Finding and checking an optimal solution to (\ref{eqn:optimalsolution}) is  difficult. Suppose that an arm set $\cup_{k\in[K]}\bar{\mathcal B}_{i,k}$ is given and that it is claimed to be the optimal solution to (\ref{eqn:optimalsolution}). It is clear that the problem of verifying that $\cup_{k\in[K]}\bar{\mathcal B}_{i,k}$ is optimal is computationally intractable; because we need to individually check  a significant fraction of the combinations of $\text{card}(\cup_{k\in[K]}\mathcal A_k)$ arms to determine which combination might give a larger expected payoff than the given arm set $\cup_{k\in[K]}\bar{\mathcal B}_{i,k}$. Since the number of combinations grows exponentially with the number of arms, the complexity of any systematic algorithm becomes impractically large.
Moreover, the expected payoff $\mathcal U_{i,k}$ depends on the unknown true state $s_{i,k}^*$, which creates yet another layer of difficulty for finding and checking an optimal solution.  

\paragraph{Variational problem}
We introduce the following notation:
$\delta_{i,k}(v) \equiv \frac{1}{2}[\max_{s_{i,k}}\pi_{i,k}(s_{i,k},v) - \min_{s_{i,k}}\pi_{i,k}(s_{i,k},v)],$
which measures the uncertainty of the acceptance probability with respect to the unknown state. 
Using this notation, we show that a variational formulation gives a practical methodology for finding the optimal strategy.
\begin{theorem}
\label{thm:reducvar}
There exist parameters $\eta_{i,k}> 0$, for $k\leq K-1$, and $\eta_{i,K}=0$ such that with high probability,
the minimizer of the following variational loss, $\forall k\in[K]$,
\begin{equation}
\label{eqn:utilitysequnm}
 \mathcal L_{i,k}^\dagger[\mathcal B_{i,k}]  = \sum_{j\in\mathcal B_{i,k}}\left(v_j+e_{ij}\right) [\eta_{i,k}\delta_{i,k}(v_j)-\pi_{i,k}(s^*_{i,k},v_j)]  + \gamma_i\max\{\mathcal N_{i,k}(\mathcal B_{i,k})-q_i,\ 0\},
\end{equation}
gives a maximizer of the total expected payoff $\sum_{k=1}^K \mathcal U_{i,k}[\mathcal B_{i,k}]$. Here the expected payoff $\mathcal U_{i,k}[\mathcal B_{i,k}]$ is given in (\ref{eqn:totalutility}), and $ \mathcal B_{i,k}\subseteq \{\mathcal A_k\setminus \cup_{l\leq k-1}\mathcal B_{i,l}\}$ for any $k\in[K]$.
\end{theorem}
\noindent
We make four remarks regarding this theorem. First, the parameter $\eta_{i,k}\geq 0$ in (\ref{eqn:utilitysequnm}) is induced by the \emph{hierarchical structure} 
in the sense that the arms available at subsequent stages are worse than the current ones; see Appendix \ref{sec:profofreducvar}. Hence, each agent prefers arms with a stable acceptance probability, and for which $\eta_{i,k}$ controls the penalty on 
the uncertainty.
Second, $\eta_{i,k}$ serves as a regularization parameter in the optimization (\ref{eqn:utilitysequnm}) for the uncertainty measure $\delta_{i,k}$. In practice, we may choose a large value of $\eta_{i,k}$ if the agents’ competition is tense, as the arms available at subsequent stages are much worse than the current ones.
Third, we note that the multi-stage decentralized matching problem is different from the multi-armed bandit problem \cite{bubeck2012, liu2019, liu2020bandit}.
A bandit problem is a sequential allocation problem in which an environment repeatedly provides an agent with a fixed  set of arms. 
Although similar in that it involves sequential decision making under limited information,  the multi-stage matching market involves multiple agents competing for arms. An arm exits the market once it accepts an offer.
The competition induces a hierarchical structure which makes  the optimization in (\ref{eqn:utilitysequnm}) different from the optimization in multi-armed bandits.
Finally, there exists a fundamental difference between the multi-stage matching when $K>1$ and the single-stage matching when $K=1$. In particular,  when $K>1$, the optimization (\ref{eqn:utilitysequnm}) has a regularization term $\eta_{i,k}\delta_{i,k}(v_j)>0$ on the uncertainty of the acceptance probability. 
In contrast, this term vanishes when $K=1$ as $\eta_{i,K}=0$.
As a result, the optimal strategy in multi-stage matching in Section \ref{sec:LUBcdm} and its economic consequences in Section \ref{sec:stregicuncertainty} are distinct from those in single-stage matching \cite{dai2020learning}.

\paragraph{Greedy strategy}
\label{sec:near-optimal}
Although the variational problem in (\ref{eqn:utilitysequnm})  requires only stage-wise optimization and can be solved sequentially for each $k\in[K]$, the finding and checking of an optimal solution is still computationally intractable. This is because we need to individually check  a significant fraction of the combinations of $\text{card}(\mathcal A_k\setminus \cup_{l\leq k-1}\mathcal B_{i,l})$ arms at each stage $k\in[K]$ to determine the optimal solution for (\ref{eqn:utilitysequnm}). The number of combinations grows exponentially with $\text{card}(\mathcal A_k\setminus \cup_{l\leq k-1}\mathcal B_{i,l})$ for $k\in[K]$.

We propose a greedy algorithm that gives an approximate solution to the optimization problem in (\ref{eqn:utilitysequnm}).
Suppose the true state is fixed at $s_{i,k}^*=s_{i,k}$.
We refer to $(v_j+e_{ij})[\pi_{i,k}(s_{i,k},v_j) - \eta_{i,k}\delta_{i,k}(v_j)]$ as arm $A_j$'s \emph{variational expected utility}. 
For each $A_j\in \{\mathcal A_k\setminus \cup_{l\leq k-1}\mathcal B_{i,l}\}$, the greedy algorithm computes the variational expected utility per unit of acceptance probability, that is,
\begin{equation*}
    r(A_j) \equiv (v_j+e_{ij})[\pi_{i,k}(s_{i,k},v_j) - \eta_{i,k}\delta_{i,k}(v_j)]/\pi_{i,k}(s_{i,k},v_j).
\end{equation*}
Then the algorithm ranks arms according to its associated value of $r$ so that $r_{(1)}\geq r_{(2)}\cdots\geq r_{(\text{card}(\mathcal A_k\setminus \cup_{l\leq k-1}\mathcal B_{i,l}))}$.  Starting with the first arm corresponding to $r_{(1)}$ and continuing in order, the algorithm selects the arm if its variational expected utility is larger than the expected penalty of exceeding the quota.
This algorithm terminates when it arrives at a cutoff value of $r$. Then only arms whose associated $r$ value are better than or equal to the cutoff are selected for agent $P_i$ to pull at stage $k\in[K]$.
We present the formalized cutoff $r=r_*$ in Appendix \ref{subsec:optofdcb}. 
Then using the greedy algorithm, agent $P_i$ pulls arms from the following set,
\begin{equation}
\label{eqn:greedy}
\widehat{\mathcal B}_{i,k}(s_{i,k})=\left.\left\{j \ \right|  \text{$A_j\in \{\mathcal A_k\setminus \cup_{l\leq k-1}\mathcal B_{i,l}\}$ satisfying } r(A_j)\geq r_*\right\}.
\end{equation} 

\begin{theorem}
\label{thm:expoptstrategy}
Suppose the true state is fixed at $s_{i,k}^*=s_{i,k}$.
The arm set $\widehat{\mathcal B}_{i,k}(s_{i,k})$ in (\ref{eqn:greedy}) is near-optimal as its loss satisfies 
\begin{equation*}
\min_{\mathcal B_{i,k}\subseteq \{\mathcal A_k\setminus \cup_{l\leq k-1}\mathcal B_{i,l}\}}\mathcal L_i^\dagger[\mathcal B_{i,k}]\leq \mathcal L_i^\dagger[\widehat{\mathcal B}_{i,k}(s_{i,k})] \leq \min_{\mathcal B_{i,k}\subseteq \{\mathcal A_k\setminus \cup_{l\leq k-1}\mathcal B_{i,l}\}}\mathcal L_i^\dagger[\mathcal B_{i,k}] + \text{UE}^\dagger,
\end{equation*}
where the loss function $\mathcal L_i^\dagger$ is defined in (\ref{eqn:utilitysequnm}). The quantity $\text{UE}^\dagger \geq 0 $ and it equals $0$ if there is a continuum of arms and $\pi_{i,k}(\cdot,v)$ is continuous in $v$.
\end{theorem}


\subsection{A two-step learning algorithm}
\label{sec:LUBcdm}
Since the true state and the acceptance probability are unknown a priori in practice, the greedy strategy in (\ref{eqn:greedy}) is unknown a priori to the agent $P_i$.
We propose a two-step algorithm to learning the greedy strategy by using historical data and statistical machine learning methods. 
The two-step algorithm is built upon the concepts of \emph{lower uncertainty bound} (LUB) and \emph{calibrated decentralized matching} (CDM) \cite{dai2020learning}.
In the first step, we compute an estimated expected utility of each arm and its lower uncertainty bound. 
Many machine learning methods can be applied here for the modeling of historical data.
In the second step, we calibrate the state parameter in a  data-driven approach that takes the opportunity cost and penalty for exceeding the quota into account.
Based on the calibrated state, an agent selects arms with the largest lower uncertainty bounds of the expected utility. 
The key idea  is to select arms which have large expected utility or little uncertainty in the expected utility.

\paragraph{Step 1: Lower uncertainty bound}

Let $\mathcal A^t = \{A_1^t,A_2^t,\ldots,A_{n^t}^t\}$ be the arm set at  $t\in[T]\equiv\{1,\ldots,T\}$. Let $s_{i,k}^t$ be the state of agent $P_i$ at stage $k$ and time $t$. The state $s_{i,k}^t$ is \emph{unknown} until the next stage or the next time point, and the state $s_{i,k}^t$  varies over time. For instance,  the yield rate  of a college may change over the years.  
 For any arm $A_j^t\in\mathcal A^t$, there are an associated pair of the  score and fit values $(v_j^{t}, e_{ij}^{t})$ obtained from (\ref{eqn:defofutility}), where $ i\in[m], j\in[n^t].$
Let $(v_j^t,e_{ij}^t)$ denote the attributes of arm $A_j^t$. 
Define the set
$\mathcal B^t_{i ,k}= \{j \ | \  P_i\text{ pulls arm } A_j^t \text{ at time } t \text{ and step } k, 1\leq j\leq n^t\}$,
where $\text{card}(\mathcal B^t_{i,k}) = n_{i,k}^t\leq n^t$.
For any $j\in\mathcal B^t_{i,k}$,  the outcome that $P_i$ observes is whether an arm $A_j^t$ accepted $P_i$, that is, $y_{ij}^t  = \mathbf 1\{A_j^t \text{ accepts } P_i\}$. 
We want to estimate $\pi_{i,k}$ based on the historical data, $\mathcal D=\{(s_{i,k}^t,v_j^t,e_{ij}^t,y_{ij}^t):i\in[m]; j\in \cup_{k=1}^K\mathcal B_{i,k}^t; t\in[T]\}$.

A wide range of machine learning methods, e.g.,  reproducing kernel methods, random forests, or neural networks, can be applied here to learn $\pi_{i,k}$ (cf. \cite{hastie2009elements}). For concreteness, we consider a penalized estimator in RKHS. 
Let the log odds ratio 
$f_{i,k}(s_{i,k},v) = \log\{\pi_{i,k}(s_{i,k},v)/[1-\pi_{i,k}(s_{i,k},v)]\}$,
which is assumed to reside in an RKHS  $\mathcal H_{\mathcal K_{i,k}}$ with the kernel $\mathcal K_{i,k}$. 
Then we solve for $\widehat{f}_{i,k} \in\mathcal H_{\mathcal K_{i,k}}$ that minimizes the objective function:
\begin{equation*}
\label{eqn:minklr}
\sum_{t=1}^T\frac{1}{n_{i,k}^t}\sum_{j\in \mathcal B_{i,k}^t}\left[-y_{ij}^tf_{i,k}(s_{i,k}^t,v^t_j)+\log\left(1+\exp\left(f_{i,k}(s_{i,k}^t,v^t_j)\right)\right)\right]+\lambda_{i,k} \|f_{i,k}\|_{\mathcal H_{\mathcal K_{i,k}}}^2,
\end{equation*}
where $\lambda_{i,k}\geq 0$ is a tuning parameter. 
Consider  the tensor product structure of $\mathcal H_{\mathcal K_{i,k}}$, where  $\mathcal K_{i,k}((s_i,v),(s_i',v')) = \mathcal K_{i,k}^s(s_i,s_i')\mathcal K_{i,k}^v(v,v')$ with some kernel functions $\mathcal K_{i,k}^s$ and $\mathcal K_{i,k}^v$ \cite{wahbawang1995}.
It is known that $\widehat{f}_{i,k}$ is minimax rate-optimal  and satisfies $\E[(\widehat{f}_{i,k} - f_{i,k})^2]\leq c_f [T(\log T)^{-1}]^{-2r/(2r+1)}$ for any $i\in[m]$  (cf. \cite{dai2020learning}).
Here, $c_f>0$ is a constant independent of $T$, and
$r\geq 1$ denotes the order of smoothness. 
The value of learning from historical data is particularly significant when a new arm is introduced into the problem.  
Let  $\mathcal A^{T+1} = \{A_{1},\ldots,A_{n}\}$ be the new arm set at time $T+1$, where  $A_j$ has  attributes obtained from (\ref{eqn:defofutility}). Then  the probability that $A_j$ accepts  $P_i$ at stage $k$ is estimated by
$\widehat{\pi}_{i,k}(s_{i,k},v_j) =\{1+\exp[-\widehat{f}_{i,k}(s_{i,k},v_j)]\}^{-1}$.
The expected utility of $A_j$ is
$\widehat{\pi}_{i,k}(s_{i,k},v_j)(v_j+e_{ij})$ for any $j\in[n]$.
Finally, we construct a lower uncertainty bound for  $\pi_{i,k}(s_{i,k},v_j)$ as,
\begin{equation}
\label{eqn:defofupconf}
\widehat{\pi}_{i,k}^{\text{L}}(s_{i,k},v_j)  =\begin{cases}
\widehat{\pi}_{i,k}(s_{i,k},v_j) - \eta_{i,k}\widehat{\delta}_{i,k}(v_j),  \\
\quad\quad\quad\text{if } v_j\in[\min\{v_j^t  \ |\ j\in\cup_{t=1}^T \mathcal B_i^t\},  \max\{v_j^t \ |\ j\in\cup_{t=1}^T \mathcal B_i^t\}];\\
1, \quad\quad \text{o.w.},
\end{cases}
\end{equation}
where $\widehat{\delta}_{i,k}(v_j) = \frac{1}{2}[\max_{s_{i,k}}\widehat{\pi}_{i,k}(s_{i,k},v_j) - \min_{s_{i,k}}\widehat{\pi}_{i,k}(s_{i,k},v_j)]$. The parameter $\eta_{i,k}\geq 0$ is defined in (\ref{eqn:utilitysequnm}).
Note that (\ref{eqn:defofupconf}) assigns  probability one to  arms with scores that agent $P_i$ has never pulled. Hence it encourages the exploration of previously untried arms.
A \emph{lower uncertainty bound} for the expected utility is then given by $\widehat{\pi}^{\text{L}}_{i,k}(s_{i,k},v_j)(v_j+e_{ij})$ for any $j\in[n]$.

The prediction of match compatibility is also possible in another direction that an arm $A_j$ can also learn how much an agent $P_i$ may like itself by predicting the probability that $A_j$ can be pulled by $P_i$. 
The arms would make the decisions based on the prediction that if they have a realistic potential of being pulled by a better agent. This feature also distinguishes the two-sided matching platform from a one-sided recommendation engine that only considers which arms an agent may like, but not which arms may also like the agent in return. 

\paragraph{Step 2: Calibrated decentralized matching}
\label{sec:calibration}
Since the true state $s_{i,k}^*$ is unknown in practice, a natural question is how to calibrate the state parameter $s_{i,k}$ in (\ref{eqn:greedy}).
Consider the average-case loss, $\E_{s_{i,k}^*}\{\mathcal L_{i,k}^\dagger[\widehat{\mathcal B}_{i,k}(s_{i,k})]\}$, where the loss $\mathcal L_{i,k}^\dagger$ is defined in (\ref{eqn:utilitysequnm}).
Define the \emph{marginal set} as
$\partial \widehat{\mathcal B}_{i,k}(s_{i,k}) \equiv \lim_{\delta_s\to 0_+} \{\widehat{\mathcal B}_{i,k}(s_{i,k}-\delta_s)\setminus \widehat{\mathcal B}_{i,k}(s_{i,k})\}$.
Hence $\partial \widehat{\mathcal B}_{i,k}(s_{i,k})$ represents the change of $\widehat{\mathcal B}_{i,k}(s_{i,k})$ with a perturbation of $s_{i,k}$.
\begin{theorem}
\label{thm:optimalstatesseq}
The average-case loss
$\E_{s_{i,k}^*}\{\mathcal L_{i,k}^\dagger[\widehat{\mathcal B}_{i,k}(s_{i,k})]\}$ is minimized if $s_{i,k}\in(0,1)$ is chosen as the solution to
\begin{equation}
\label{eqn:meancalib}
\begin{aligned}
& \mathbb P(s_{i,k}^*\neq s_{i,k})\sum_{j\in\partial\widehat{\mathcal B}_{i,k}(s_{i,k})}(v_j+e_{ij})\E_{s_{i,k}^*}\left[\pi_{i,k}(s_{i,k}^*,v_j)- \eta_{i,k}\delta_{i,k}(v_j)\ | \ s_{i,k}^*\neq s_{i,k}\right]\\
& \quad\quad\quad\quad  = \gamma_i[1-F_{s_{i,k}^*}(s_{i,k})]\sum_{j\in\partial\widehat{\mathcal B}_{i,k}(s_{i,k})}\E_{s_{i,k}^*}[\pi_{i,k}(s_{i,k}^*,v_j)\ |\ s_{i,k}<s_{i,k}^*\leq 1],
\end{aligned}
\end{equation}
where $F_{s_{i,k}^*}$ is the cumulative distribution function of $s_{i,k}^*\in[0,1]$.
\end{theorem}
\noindent
The key idea of (\ref{eqn:meancalib}) is to balance the trade-off between opportunity cost and penalty for exceeding the quota.  If (\ref{eqn:meancalib}) has more than one solution, then $s_{i,k}$ is chosen as the largest one. If the distribution $F_{s_{i,k}^*}$ has discrete support, the objective in  Theorem \ref{thm:optimalstatesseq} needs to be changed as follows:  choosing the minimal $s_{i,k}\in[0,1]$ such that the left side of (\ref{eqn:meancalib}) is not less than the right side of (\ref{eqn:meancalib}), where the search of $s_{i,k}$ starts from the maximum value in the support and decreases to the minimal value. 
Moreover, instead of the average-case loss in Theorem \ref{thm:optimalstatesseq}, we can also perform the calibration under the worst-case loss, which is discussed in Appendix \ref{sec:calminvarcdm}.

\begin{algorithm}[t!]
\caption{The two-step  algorithm for multi-stage decentralized matching} 
\begin{algorithmic}[1]
\STATE \textbf{Inputs}: Historical data for an agent $P_i$: $\{(s_{i,k}^t,v_j^t,e_{ij}^t,y_{ij}^t): j\in \mathcal B_{i,k}^t; t=1,2,\ldots,T\}$; New arm set $\mathcal A^{T+1}$ at time $T+1$, where the arms have attributes $\{(v_j,e_{ij}):j\in[n]\}$; Penalty $\gamma_i$ for exceeding the quota. Regularization parameter $\eta_{i,k}\geq 0$.
\FOR{stage $k=1,2,\ldots,K$}
    \STATE 	Construct the lower uncertainty bound  $\widehat{\pi}_{i,k}^{\text{L}}(s_{i,k},v_j)$ by (\ref{eqn:defofupconf})
    \STATE Estimate the distribution $F_{s_{i,k}^*}(\cdot)$ by the kernel density method \cite{silverman1986}
    \STATE Calibrate the state $s_{i,k}$ according to Theorem \ref{thm:optimalstatesseq}
    \STATE Determine the  arm set $\widehat{\mathcal B}^{\text{L}}_{i,k}(s_{i,k})$ in (\ref{eqn:bbL})
    \STATE Calculate the remaining quota: $q_{i}-\text{card}(\cup_{l\leq k-1}\mathcal C_{i,l})$ and the available arms
\ENDFOR
\STATE \textbf{Outputs}: The arm set $\widehat{\mathcal B}^{\text{L}}_{i,k}(s_{i,k})$ for agent $P_i$ at each stages.
\end{algorithmic} 
\label{alg:decentralizedcdmseq}
\end{algorithm}

\paragraph{Summary of the two-step algorithm}
Using (\ref{eqn:defofupconf}) and (\ref{eqn:meancalib}), we can obtain the cutoff estimate $\widehat{r}_*$ and calibrated state $s_{i,k}$, which suggests agent $P_i$ to pull arms from the following set at stage $k$:
\begin{equation}
\label{eqn:bbL}
\widehat{\mathcal B}^{\text{L}}_{i,k}(s_{i,k})=\left.\left\{j \ \right|   \text{$A_j\in\{\mathcal A_k^{T+1}\setminus \cup_{l\leq k-1}\mathcal B_{i,l}\}$ satisfying }r(A_j)\geq \widehat{r}_*\right\}.
\end{equation}
Here $\mathcal A_k^{T+1}$ is the set of arms that are available at stage $k$ of time $T+1$.
Due to the minimax optimality of $\widehat{f}_{i,k}$, we have the consistency result that
$\widehat{\mathcal B}_{i,k}^\text{L}(s_{i,k})\to \widehat{\mathcal B}_{i,k}(s_{i,k})$ as $T\to\infty$,
where the set $\widehat{\mathcal B}_{i,k}(s_{i,k})$ is defined in (\ref{eqn:greedy}). 
We summarize the above two-step algorithm in Algorithm \ref{alg:decentralizedcdmseq}.
We also remark that
although the negligible application costs is assumed in Section \ref{sec:model}, Algorithm \ref{alg:decentralizedcdmseq} is applicable to non-negligible application costs, in which different agents (i.e., colleges) would have different sets of available arms (i.e., student applicants).


\section{Strategic Behavior and Economic Implications}
\label{sec:stregicuncertainty}
Agents in a multi-stage decentralized matching markets cannot observe other agents' quotas or the choices of the arms that accept other agents. Each agent only observes the arms that are left in the  market at each stage. Theorem \ref{thm:reducvar} implies that agents prefer arms with stable acceptance probability.  This preference lead to strategic behavior on the part of the agents as follows.
Define the \emph{uncertainty level} as  the uncertainty measure $\delta_{i,k}(v)$ in Section \ref{sec:modifiedutility} relative to the acceptance probability $\pi_{i,k}(s_{i,k},v)$. That is,
\begin{equation}
\label{eqn:uncertaintylevel}
 \text{uncertainty level} \equiv  \delta_{i,k}(v)/\pi_{i,k}(s_{i,k},v).  
\end{equation}
We show in Appendix \ref{subsec:optofdcb} that the cutoff $r_*$ in (\ref{eqn:greedy}) is strictly increasing in the uncertainty level for any $v\in[0,1]$ and $k\leq K-1$, which implies that an agent favors arms with a low uncertainty level. Hence, an agent's \emph{strategic behavior} in this market is to strategically select arms with a low uncertainty level. 
We now study the implications of such strategic behavior on  fairness and welfare. 

\paragraph{No justified envy}
The fairness studied here is defined in terms of \emph{no justified envy} \cite{Abdulkadiroglu2003, balinski1999}. Specifically, 
an arm $A_j$ has justified envy if, at a stage $k\in[K]$, $A_j$ prefers an agent $P_{i'}$ to another agent $P_i$ that pulls $A_j$, even though $P_{i'}$ pulls an arm $A_{j'}$ which ranks below $A_j$ according to the true preference of $P_{i'}$.
We define a multi-stage matching procedure to be \emph{fair} if there is no arm having justified envy at any stage.

\begin{proposition}
\label{prop:fallthroughcracksstrategy}
The probability that an arm has justified envy is strictly increasing in the arm's uncertainty level defined in (\ref{eqn:uncertaintylevel}). 
\end{proposition}
\noindent
The fairness issue has been noted in practical multi-stage matching markets. For example, candidates in job markets may ``fall through the cracks"---an employer  that  values  a  candidate  highly  perceives  that  the  candidate  is  unlikely  to  accept  the  job offer  and  hence  declines  to  conduct  an  interview  with  the candidate; hence,  candidates may have justified envy \cite{coles2010job}. 
Besides our ex-ante definition of no justified envy, there are other choices of no justified envy, including ex-post definition, which could lead to a different set of technical results \cite{freeman2020best}.

\paragraph{Fairness vs. welfare trade-off}
We note that by Theorem \ref{thm:reducvar}, an agent has increased expected payoff under $\eta_{i,k}>0$ than under $\eta_{i,k}=0$ for all stages $k\leq K-1$. 
Define the number of arms with justified envy to be the \emph{level of justified envy} of the matching outcome. 
Then if the level of justified envy is zero, the matching outcome is fair for arms.

\begin{proposition}
\label{thm:fairnessandeta}
The level of justified envy is strictly increasing in $\eta_{i,k}\geq 0$.
\end{proposition}
This proposition implies a trade-off between welfare and fairness since both the level of justified envy and welfare increase when changing $\eta_{i,k}=0$ to $\eta_{i,k}>0$.
We give an example of two-stage decentralized matching, that is, $K=2$. Such two-stage matching is typical in college admissions, which may include regular admissions and waiting-list admissions. By Theorems \ref{thm:reducvar} and  \ref{thm:expoptstrategy}, agents in the first stage would strategically pull arms with low uncertainty levels by taking $\eta_{i,1}>0$.
In this way, agents would reduce head-on competition.
Next, agents in the second stage would act according to their true preferences and pull available arms with top latent utilities by taking $\eta_{i,2}=0$.  Theorem \ref{thm:reducvar} shows that agents' strategic behavior in the first stage increases the welfare compared to acting according to their true preferences, whereas in the second stage, agents acting according to their true preferences suffices.
Proposition \ref{thm:fairnessandeta} shows that agents' strategic behavior in the first stage results in increased welfare, but at the cost of arms' fairness.

\paragraph{Comparison with single-stage matching markets}
Different from multi-stage matching markets, the optimal strategy in single-stage matching gives a fair outcome for arms \cite{dai2020learning}. However, we show 
that agents are better off in multi-stage markets compared to single-stage markets.
\begin{proposition}
\label{prof:benefitofmulstamat}
Agents have improved welfare  under multi-stage decentralized matching than under single-stage decentralized matching. 
\end{proposition}
We provide an empirical example in Appendix \ref{sec:multivssinglematching} to illustrate the gap between multi-stage welfare and single-stage welfare.


\paragraph{Comparison with centralized matching markets}
Many centralized matching markets are implemented by employing the celebrated deferred acceptance (DA) algorithm \cite{gale1962}; see examples in \cite{Abdulkadiroglu2003, roth1984}. 
In the arm-proposing version of DA (e.g., student-proposing in college admissions), agents and arms report their ordinal preferences to a clearinghouse,  which simulates the following multi-stage procedure. Every arm shows its interest to the most preferred agent that has not yet rejected it at each stage. 
Every agent tentatively pulls the most preferred arms up to its quota limit and permanently rejects the remaining arms that have indicated their interest to the agent. 
Once the process terminates, 
each arm is assigned to the agent that has tentatively pulled it or otherwise remains unmatched. 
The multi-stage decentralized matching is different from DA in practice, mainly due to the acceptance is not tentative (i.e., non-deferrable) in decentralized matching. Moreover, there is usually a  restriction on the number of stages in decentralized matching due to the time cost at each stage of multi-stage decentralized matching is not negligible. 
We show in a numerical example of Appendix \ref{sec:agentbetteroff} that some agents are better off in decentralized markets than centralized markets. This finding gives a partial explanation of the prevalence of decentralized college admissions in many countries. 

\section{Numerical Studies}
\label{sec:simulation}

In this section we demonstrate aspects of the theoretical predictions through a simulation and a real data application in college admissions. We provide extensive numerical comparisons of Algorithm \ref{alg:decentralizedcdmseq} with other methods in  Appendix.
We also give additional real data analysis in Appendix.
The total computing hour is within one hour in personal laptop with Intel Core i5.

\paragraph{Simulated graduate school admissions}
Consider $50$ graduate schools from three tiers of colleges: five top colleges $\{P_1,\ldots,P_5\}$, ten good colleges $\{P_6,\ldots,P_{15}\}$, and $35$ other colleges $\{P_{16},\ldots,P_{50}\}$. Each has the same quota $q=5$ and penalty $\gamma=2.5$. The simulation generates students' preferences with ten different states $\{s_1,\ldots,s_{10}\}\subset[0,1]$. For any state, students' preferences for colleges from the same tier are random. However,  students prefer top colleges to good colleges, and the other colleges are the least favorite.
The random preferences depend on the state due to colleges' uncertain reputation and popularity in the current year.  
We consider varying numbers of students $\{250,260,270,280,290,300\}$.
For each size of students, there are ten students having score $v_j$ chosen uniformly and i.i.d.\ from $[0.9,1]$ and $100$ students having score $v_j$ i.i.d. uniformly chosen from $[0.7,0.9)$. The rest of the students have score $v_j$  randomly chosen from $[0,0.7)$. The fits $e_{ij}$ for all college-student pairs are drawn uniformly and i.i.d.\ from $[0,1]$. 

\begin{figure}[ht]
\centering
\includegraphics[width=\textwidth, height=2.25in]{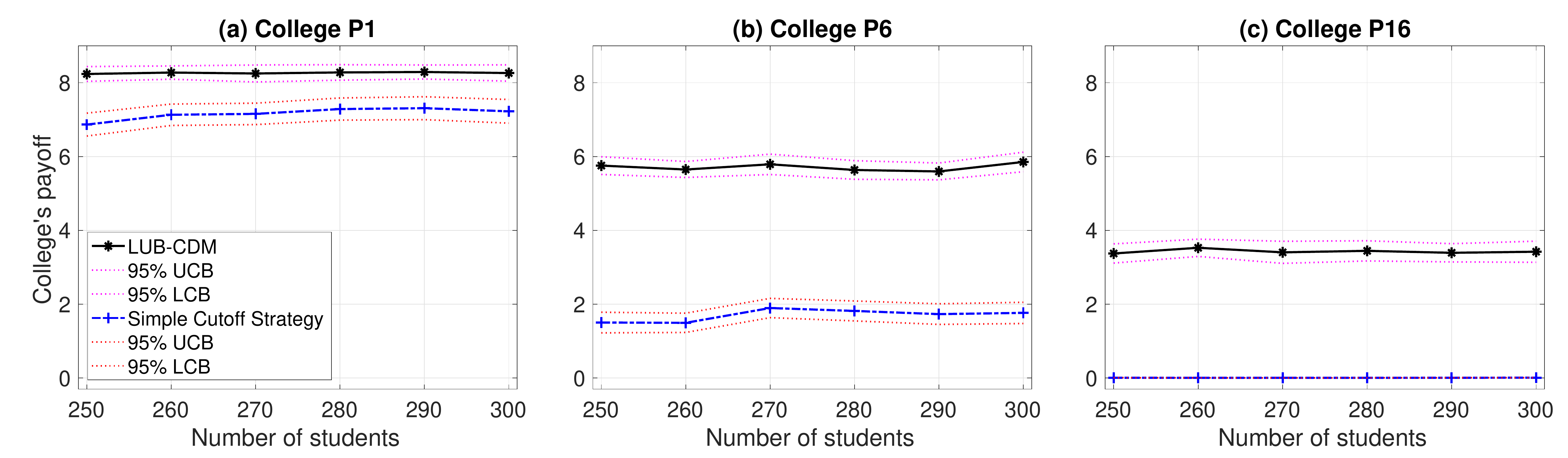}
\caption{Performance of the proposed Algorithm \ref{alg:decentralizedcdmseq} (i.e., LUB-CDM) and the Simple Cutoff Strategies with varying numbers of students. The results are averaged over $500$ data replications. (a): College $P_1$ from tier 1. (b): College $P_6$ from tier 2. (c): College $P_{16}$ from tier 3.}
\label{fig:graduateschool}
\end{figure}

We compare the college’s expected payoff achieved by the proposed Algorithm \ref{alg:decentralizedcdmseq} with the \emph{simple cutoff strategy}, where the latter method has each college choosing the most preferred students up to the remaining quota at each stage.    The training data are simulated from colleges’ random proposing by pulling a random number of arms according to the latent utilities. The training data consists of $20$ times of random proposing under each of the arms' preference structures with the two-stage admissions. This training data simulates the graduate school admissions over $20$ years.  The testing data draws a random state from $\{s_1,\ldots,s_{10}\}$ which gives the corresponding arms' preferences. Then we apply Algorithm \ref{alg:decentralizedcdmseq} with $\eta_{i,1}=0.1,\eta_{i,2} = 0$ and $\gamma_i=2.5$.
Figure \ref{fig:graduateschool} reports the averaged payoffs of three colleges $P_1,P_6$, and $P_{16}$ over $500$ data replications. Here colleges $P_1,P_6$, and $P_{16}$ belong to the three different tiers, respectively.
In Figure \ref{fig:graduateschool}, all  colleges except $P_1$ use Algorithm \ref{alg:decentralizedcdmseq} while $P_1$ uses one of the two methods: Algorithm \ref{alg:decentralizedcdmseq}  and the simple cutoff strategy. 
It is seen that Algorithm \ref{alg:decentralizedcdmseq} gives the largest average payoffs for all of $P_1,P_6$ and $P_{16}$. In particular, Algorithm \ref{alg:decentralizedcdmseq} performs significantly better for $P_6$ and $P_{16}$  compared to the simple cutoff strategy. 


\paragraph{U.S. college admissions}
We study a public data on college admissions from the \emph{New York Times} ``The Choice" blog. 
In this dataset, $37$ U.S. colleges reported their admission yields and waiting list offers for 2015--17  applicants without
personally identifiable information.
As we discussed in Section \ref{sec:model}, a college's yield is a proxy for the state $s_{i,k}$ as it indicates the college's popularity.
The set of $37$ colleges consists of liberal arts colleges, national universities, and other undergraduate programs.

\begin{figure}[ht]
    \centering
    \includegraphics[width=\textwidth]{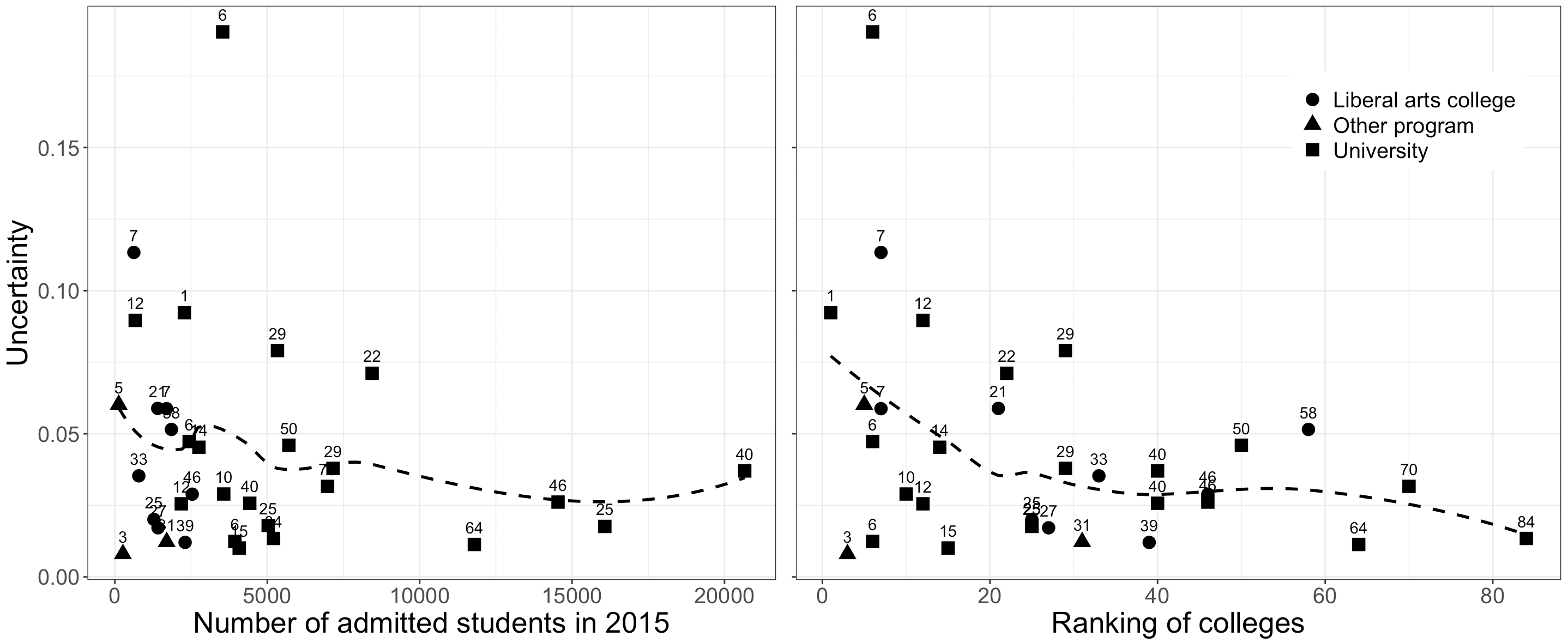}
    \caption{ Regression of  uncertainty level on the size of admitted class and the ranking, respectively. 
    Two dashed curves are fitted using smoothing splines with the tuning parameter chosen by GCV.}
    \label{fig:NYTuncertainty}
\end{figure}

We estimate the uncertainty level $\delta_{i,k}(v)\pi^{-1}_{i,k}(s_{i,k},v)$ defined in  (\ref{eqn:uncertaintylevel}) and study colleges' strategic responses. 
While conclusive evidence on the individual students' acceptance probability is difficult to obtain, we estimate the college-wise  uncertainty on the yield: $\sqrt{\text{Var}(s_{i,k})}s^{-1}_{i,k}$. Since the choice set for admitted students differs across years, the yield's uncertainty underestimates the uncertainty facing a college. 
Figure \ref{fig:NYTuncertainty} shows  that colleges' uncertainty levels are much smaller than one, which, together with Theorem \ref{prop:fallthroughcracksstrategy}, implies that students face limited unfairness. In particular, the yield uncertainty is robust to the size of admitted students; see the left plot of  Figure \ref{fig:NYTuncertainty}.
On the other hand, top-ranked national universities may have higher uncertainty levels; see the right plot of  Figure \ref{fig:NYTuncertainty}, where the outlier is the University of Chicago at the $.19$ uncertainty level. 
We verify the higher uncertainty level for top universities using the waiting list data.  We perform Fisher's exact test for the rank data on the difference of rates of accepted waiting list students to total enrolled students over 2015--16.
This statistic reflects the uncertainty on both the regular admission yield and  the wait-listed students' quality. We reject the null hypothesis that the uncertainty of acceptance is the same for all national universities at the $.05$ significance level. 
The higher uncertainty for top-ranked national universities may arise due to the intense competition. Those universities are better off by employing strategic admission to reduce the enrollment uncertainty. 
This result implies that students are more likely to experience unfairness when applying for top national universities.   

\section{Conclusion}
\label{sec:discussion}

This paper develops a nonparametric statistical model to learn optimal strategies in multi-stage decentralized matching markets. 
The model provides insight into the interplay between learning and economic objectives in decentralized matching markets.
In the model, arms have uncertain preferences that depend on the unknown state of the world and competition among the agents.
We propose an algorithm, built upon the concepts of  lower uncertainty bound and calibrated decentralized matching, for learning optimal strategies using  historical data.  We find that agents act strategically in favor of arms with low uncertainty levels of acceptance. 
The strategic targeting improves an agent’s welfare but leads to unfairness for arms.
Our theory allows analytical comparisons between single-stage decentralized markets and centralized markets. 

For future directions,
it is of interest to study algorithmic strategies when agents' preferences show complementarities or indifference. These settings have important applications, as firms may demand workers that complement one another in terms of their skills and roles, or some applicants are indistinguishable to a firm.  We leave these questions for future work.

The problem of machine learning in economics has become increasingly important in many application domains. 
In this work, we aim to deepen the understanding of decentralized matching markets from a learning perspective and propose an efficient and scalable algorithm to solve optimal strategies.  
We do not foresee any negative impact to society from our work.

\section*{Acknowledgments and Disclosure of Funding}
We would like to thank the area chair and four anonymous referees for constructive suggestions that
improve the paper.
We thank Robert M. Anderson and Joel Sobel for helpful discussions. 
This work was supported in part by the Vannevar Bush Faculty Fellowship program under grant number N00014-21-1-2941.

\bibliographystyle{acm}
{
\bibliography{match}
}

\newpage 

\appendix


\section{Supplementary Numerical Results}

\subsection{Comparison with the straightforward strategy}
\label{exp:egthreeagents}
In this example, we compare the proposed Algorithm \ref{alg:decentralizedcdmseq} with the \emph{straightforward strategy}, where the latter method pulls arms according to the latent utility defined in Eq.~\eqref{eqn:defofutility} and calibrates the state in the same way as Algorithm \ref{alg:decentralizedcdmseq}.

Suppose there are $n$ arms $\mathcal A = \{A_1,A_2,\ldots,A_n\}$ and three agents $\mathcal P=\{P_1,P_2,P_3\}$, where each agent has a quota $q<n/3$.
There are  two equally likely states: $s_a$ and $s_b$ with $s_a=1-s_b> 1/2$. All arms prefer $P_1$ and $P_2$ to $P_3$, but the arms prefer $P_3$  compared to being unmatched.  
Agents $P_1$ and $P_2$ evaluate each arm based on score $v$ and with probability $p^*\in(0,1)$, each of $P_1$ and $P_2$ finds an arm  unacceptable. 
Agent $P_3$ evaluates each arm only based on the score. 
For each state $j\in\{a,b\}$, a fraction $s_j$ of arms receives utility $u_1$ when matched to $P_1$ and utility $u_2$ when matched to $P_2$, where $u_1>u_2$ and the remaining $(1-s_j)$ of arms receive the opposite utilities. 
Hence, $P_1$ is more popular under the state $s_a$ and $P_2$ is more popular under the state $s_b$. In each state, an arm gets utility $u_3$ from $P_3$, where $(1-p^*)u_1<u_3<u_1$. This condition implies that an arm is better off by accepting $P_3$ than waiting for $P_1$ or $P_2$.  
We consider a two-stage matching, where at the first stage,  each agent pulls a set of arms and wait-lists other arms. An arm  pulled by an agent must accept or reject the agent immediately. 
\renewcommand{\theproposition}{\thesection.\arabic{proposition}}
\begin{proposition}
\label{thm:explimprovs}
Agent $P_1$ is better off by using Algorithm \ref{alg:decentralizedcdmseq}  than using the straightforward strategy, where the expected payoff is improved by $O(\eta_{1,1})$. Here $\eta_{1,1}$ is the regularization parameter defined in Theorem \ref{thm:reducvar}.
\end{proposition}

\begin{figure}[ht]
    \centering
    \includegraphics[width=\textwidth]{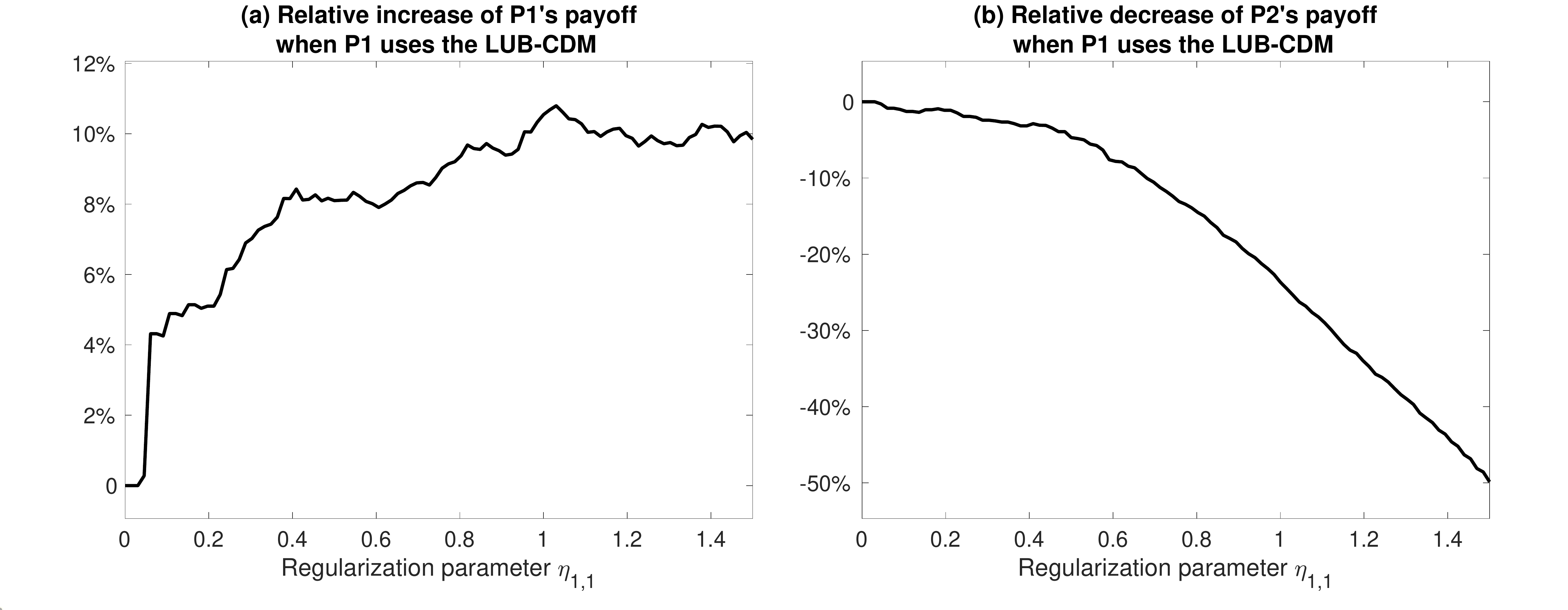}
    \caption{Comparison of the proposed Algorithm \ref{alg:decentralizedcdmseq} (i.e., LUB-CDM) and the straightforward strategy. The results are averaged over $500$ data replications.  (a) The relative increase of $P_1$'s payoffs when $P_1$ changes from the straightforward strategy to the LUB-CDM, where the improvement is $O(\eta_{1,1})$. (b) The relative decrease of $P_2$'s payoffs when $P_1$ changes from  the straightforward strategy to the LUB-CDM. }
    \label{fig:eg4}
\end{figure}

To illustrate the improvement, we consider the states $s_a=0.6, s_b=0.4$, the number of arms $n = 100$, the quota $q=10$, the utilities $u_1=1,u_2=0.9,u_3=0.8$, and the probability $p^*=0.3$. Suppose that the score $v$ follows a deterministic uniform design points $\{1.05,1.1, 1.15, \ldots,2.95,3\} \subset [1,3]$. The penalties of exceeding the quota are $\gamma_1 =\gamma_2=\gamma_3=5$. 
We compare the proposed Algorithm \ref{alg:decentralizedcdmseq} (i.e., LUB-CDM) with the straightforward strategy (i.e., CDM). The latter method is a straightforward strategy as it pulls arms according to the latent utilities in Eq.~\eqref{eqn:defofutility} without strategic behaviors.
Figure \ref{fig:eg4} reports $P_1$'s and $P_2$'s relative changes in payoffs, when $P_1$ changes from using the CDM to using the LUB-CDM.
The results are averaged over $500$  data replications. 
Here $P_1$ using the LUB-CDM and the CDM correspond to $\eta_{1,1}> 0$ and $\eta_{1,1}=0$, respectively. The $P_2$ uses CDM. 
It is seen the LUB-CDM improves  $P_1$'s expected payoff, where the improvement is at the cost of $P_2$'s payoff.  


\subsection{Comparison with the patient strategy}
\label{sec:adachi}
In this example, we compare the proposed Algorithm \ref{alg:decentralizedcdmseq} with the \emph{patient strategy}, where the latter method pulls arms according to the latent utility at the beginning stage but has more strategic behaviors as the matching proceeds. 
We consider a search model due to \cite{adachi2003}, which captures the search process in matching markets and builds a connection between the multi-stage decentralized matching markets and the centralized matching markets.

Suppose there are $n$ arms $\mathcal A=\{A_1,A_2,\ldots,A_n\}$ and $m$ agents $\mathcal P=\{P_1,P_2,\ldots,P_m\}$, where each agent has quota $q=1$. At each stage, each agent comes across a randomly sampled arm. 
Let $v_{\mathcal P}(i)$ and $v_{\mathcal A}(j)$ be the reservation utilities of agent $P_i$ and arm $A_j$ from staying unmatched and continuing the search.  Recall the latent utility $U_i(A_j)$ in Section \ref{sec:model}. Similarly, we define $U_j(P_i)$ as the utility that arm $A_j$ receives when matched to $P_i$. 
Let $v_{\mathcal P}(i)$ and $v_{\mathcal A}(j)$ be the reservation utilities of agent $P_i$ and arm $A_j$ from staying single and continuing the search for a match. Hence $\mathbf 1\{P_i\text{ pulls } A_j\}  = \mathbf 1\{U_i(A_j)\geq v_{\mathcal P}(i)\}$, and $\mathbf 1\{A_j\text{ accepts } P_i\}  = \mathbf 1\{U_j(P_i)\geq v_{\mathcal A}(j)\}$.
The utility that agent $P_i$ gets upon coming across arm $A_j$ is
\begin{equation*}
\begin{aligned}
    \bar{U}_i(A_j) & = U_i(A_j) \mathbf 1\{U_i(A_j)\geq v_{\mathcal P}(i)\} \mathbf 1\{U_j(P_i)\geq v_{\mathcal A}(j)\}\\
    &\quad\quad\quad +v_{\mathcal P}(i)[1-\mathbf 1\{U_i(A_j)\geq v_{\mathcal P}(i)\} \mathbf 1\{U_j(P_i)\geq v_{\mathcal A}(j)\}],
\end{aligned}
\end{equation*}
where the first term on the right-hand side is the utility from a successful match and the second term on the right-hand side is the utility when no match occurs. 
Adachi's model involves a stage discount factor $\rho>0$, where the Bellman equations for the optimal reservation values and search rules are
\renewcommand{\theequation}{\thesection.\arabic{equation}}
\begin{equation}
\label{eqn:bellman}
    v_{\mathcal P}(i) = \rho\int \bar{U}_i(A_j)dF_{\mathcal A}(j)\quad\text{and}\quad  v_{\mathcal A}(j) = \rho\int \bar{U}_j(P_i)dF_{\mathcal P}(i),
\end{equation}
where $F_{\mathcal A}$ and $F_{\mathcal P}$ are the distributions that each agent and arm came across. 
In \cite{adachi2003} the author shows that Bellman equations in Eq.~(\ref{eqn:bellman}) defines an iterative mapping that converges to the equilibrium reservation utilities $(v_{\mathcal P}^*(i),v_{\mathcal A}^*(j))$. Furthermore, as $\rho\to 1$, the Bellman equations lead to the matching outcomes that are stable in the sense of Gale and Shapley \cite{gale1962}. 

\begin{figure}[ht]
    \centering
    \includegraphics[width=\textwidth]{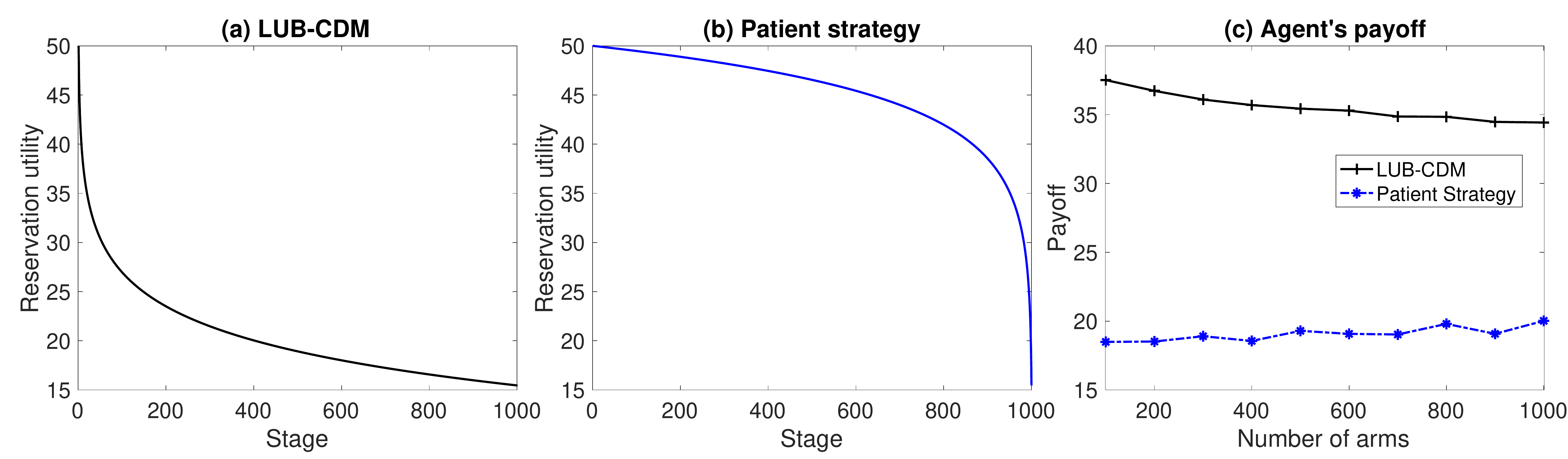}
    \caption{Performance of the proposed Algorithm \ref{alg:decentralizedcdmseq} (i.e., LUB-CDM) and the patient strategy. The results averaged over $500$ data replications. (a) $P_1$'s reservation utility under the LUB-CDM given by $50 - 5\text{log}(k)$, where the stage $k=1,\ldots,500$. (b) $P_1$'s reservation utility under the patient strategy given by $ 50 +5\text{log}(\frac{N+1-k}{N})$. 
    (c) $P_1$'s payoffs with varying number of arms. }
    \label{fig:policyeg1}
\end{figure}

Since the equilibrium reservation utilities $(v_{\mathcal P}^*(i),v_{\mathcal A}^*(j))$ are unknown in practice, agents need to learn an optimal strategy of choosing the reservation utility $v_{\mathcal P}(i)$ at different stages. We compare the proposed Algorithm \ref{alg:decentralizedcdmseq} (i.e., LUB-CDM) with the patient strategy, where the latter  is defined as the strategy with $\rho= 1$ at the beginning stage $k=1$ and decreasing $\rho$  as the matching proceeds in Eq.~(\ref{eqn:bellman}).
Note that  LUB-CDM  has less strategic behaviors as the matching proceeds.
Hence it corresponds to the case that $v_{\mathcal P}(i)$ is a convex function of the stages.  
On the other hand, the patient strategy has more strategic behaviors as the matching proceeds. Hence it
corresponds to the case that $v_{\mathcal P}(i)$ is a concave function of the stages. 
Suppose that different arms receive the same utility for matching the same agent, that is, $U_j(P_i) = U_{j'}(P_i),\forall j\neq j'$, which utility is unknown to $P_i$.  Similarly, different agents receive the same utility for matching the same arm, that is, $U_i(A_j) = U_{i'}(A_j),\forall i\neq i'$, which utility is known to $A_j$. Then $P_i$ matches with $A_j$ if the event $\{U_j(P_i)\geq U_i(A_j)\geq v_{\mathcal P}(i)\}$ holds. 
Suppose that agent $P_1$'s utility is $U_j(P_1)=40$, and $m=n\in\{100,200,\ldots,1000\}$. 
Let the reservation utility $v_{\mathcal P}(1)$ at the stage $k$ be $50 - 5\text{log}(k)$ and $ 50 +5\text{log}(\frac{N+1-k}{N})$ for the LUB-CDM and the patient strategy, respectively
; see Figure \ref{fig:policyeg1}(a) and (b). 
Figure \ref{fig:policyeg1}(c) reports $P_1$'s payoff under two methods, where the LUB-CDM outperforms the patient strategy. Therefore, the strategic behavior at early stages improves the agent's payoff in practice, which result corroborates Theorem \ref{thm:reducvar}.

\subsection{Comparison of multi-stage matching and DA}
\label{sec:agentbetteroff}

In this example, we compare the multi-stage decentralized matching with the DA algorithm \cite{gale1962}. 
Suppose there are four arms $\mathcal A=\{A_1,A_2,A_3,A_4\}$ and  three agents $\mathcal P=\{P_1,P_2,P_3\}$. Agents have varied quotas: $q_1=2$ and $q_2=q_3=1$. Arms' attributes are given by $v_1=v_2=v_3=2, v_4=1$, and $e_{13}=e_{23}=e_{32}=0,e_{12}=e_{22}= e_{31}=0.5, e_{11}=e_{21}=e_{33}=1$, $e_{14}=0.2$, $e_{24}=0.5$, $e_{34}=0.8$.
The latent utilities and arms' true preferences are shown in Table \ref{table:eg5}.
For the decentralized matching, suppose that at each stage, every agent uses the straightforward strategy by pulling its most preferred arms up to the quota. Arms accept their most preferred agent (if any) or wait until the next stage. Then the decentralized matching has the outcome $(A_1,P_1), (A_2,P_1), (A_3,P_3), (A_4,P_2)$. On the other hand, the DA algorithm gives the outcome $(A_1,P_3), (A_2,P_2), (A_3,P_1), (A_4,P_1)$, which the unique stable matching outcome.
Here both $P_1$ and $P_3$ strictly prefer the decentralized matching outcome to DA outcome. This result corroborates
the remark in Section \ref{sec:stregicuncertainty} that some agents are better off under the decentralized matching.

\begin{table}[ht]
\caption{(a) Arm's latent utilities for each agent, which corresponds to Eq.~(\ref{eqn:defofutility}). 
(b) Arms' preferences with the number indicating the arms' ranking of agents. For example, $A_1$ ranks $P_3$ first, $P_1$ second, $P_2$ third. These preferences are unknown to agents.}
\label{table:eg5}
\begin{minipage}{\columnwidth}
\centering
\begin{tabular}{ cc }   
 \textbf{\small{(a) Arm's latent utility}}  & \quad\quad\quad\quad\quad\quad  \textbf{\small{(b) Arm's preference}}  \\[0.5ex]
\begin{tabular}{ c  c c c c } 
\toprule
 & $A_1$ & $A_2$ & $A_3$  & $A_4$\\ [0.3ex]  
 \cmidrule(lr){2-2}\cmidrule(lr){3-3}\cmidrule(lr){4-4}\cmidrule(lr){5-5}
$P_1$ & 3 & 2.5 & 2 & 1.2 \\ [0.3ex]  
$P_2$ & 3 & 2.5 & 2 & 1.5\\ [0.3ex]  
$P_3$ & 2.5 & 2 & 3 & 1.8\\ [0.3ex]  
\bottomrule
\end{tabular} &  
\quad\quad\quad\quad\quad\quad
\begin{tabular}{ c  c c c  c } 
\toprule
 & $A_1$ & $A_2$ & $A_3$ & $A_4$ \\ [0.3ex]  
 \cmidrule(lr){2-2}\cmidrule(lr){3-3}\cmidrule(lr){4-4}\cmidrule(lr){5-5}
$P_1$ & 2 & 2 & 1 & 1 \\ [0.3ex]  
$P_2$ & 3 & 1 & 3 & 2 \\ [0.3ex]  
$P_3$ & 1 & 3 & 2 & 3 \\ [0.3ex]  
\bottomrule
\end{tabular}   
\end{tabular}
\end{minipage}
\end{table}

Second, we study the incentive of agents in the multi-stage decentralized  matching. 
We show that it is not a dominant strategy  for each agent to use the straightforward strategy by pulling arms according to the latent utility. For example, consider the preferences  in Table \ref{table:eg5}.  If $P_2$ skips over $A_1$ and firstly pulls $A_2$, and other agents pull their most preferred arms up to their quotas. Then the decentralized matching has the outcome $(A_1,P_1), (A_2,P_2), (A_3,P_3), (A_4,P_1)$, where $P_2$ is strictly better off  compared to the outcome when $P_2$  firstly pulls $A_1$.

\begin{table}[ht]
\caption{(a) Arm's latent utilities for each agent. (b) Arms' preferences with the number indicating the arms' ranking of agents. For example, $A_1$ ranks $P_4$ first, $P_1$ second, $P_3$ third, $P_2$ fourth.}
\label{table:eg6}
\begin{minipage}{\columnwidth}
\centering
\begin{tabular}{ cc }   
 \textbf{\small{(a) Arm's latent utility}}  & \quad\quad\quad\quad\quad\quad  \textbf{\small{(b) Arm's preference}}  \\[0.5ex]
\begin{tabular}{ c  c c c c } 
\toprule
 & $A_1$ & $A_2$ & $A_3$  & $A_4$\\ [0.3ex]  
 \cmidrule(lr){2-2}\cmidrule(lr){3-3}\cmidrule(lr){4-4}\cmidrule(lr){5-5}
$P_1$ & 3 & 2 & 2.6 & 2.3 \\ [0.3ex]  
$P_2$ & 2 & 2.6 & 3 & 2.3\\ [0.3ex]  
$P_3$ & 2.3 & 2 & 3 & 2.6\\ [0.3ex]  
$P_3$ & 2 & 2.3 & 2.6 & 3\\ [0.3ex]  
\bottomrule
\end{tabular} &  
\quad\quad\quad\quad\quad\quad
\begin{tabular}{ c  c c c  c } 
\toprule
 & $P_1$ & $P_2$ & $P_3$ & $P_4$ \\ [0.3ex]  
 \cmidrule(lr){2-2}\cmidrule(lr){3-3}\cmidrule(lr){4-4}\cmidrule(lr){5-5}
$A_1$ & 2 & 4 & 3 & 1 \\ [0.3ex]  
$A_2$ & 4 & 2 & 1 & 3 \\ [0.3ex]  
$A_3$ & 1 & 3 & 4 & 2 \\ [0.3ex]  
$A_3$ & 3 & 1 & 2 & 4 \\ [0.3ex]  
\bottomrule
\end{tabular}   
\end{tabular}
\end{minipage}
\end{table}

Finally, we show that arms can also be better off  if they are strategic in multi-stage decentralized matching.
Suppose there are four agents and four arms, and each agent  has a quota one. The latent utilities and arms' true preferences are given in Table \ref{table:eg6}. 
When agents and arms are not strategic, the decentralized matching has the outcome $(A_1,P_1),(A_2,P_3),(A_3,P_2),(A_4,P_4)$. 
However, suppose arms are strategic, where $A_4$ rejects $P_4$ as $P_4$ is $A_4$'s least favorite agent and $A_4$ believes the coming agent will not be worse. The outcome becomes $(A_1,P_1),(A_2,P_4),(A_3,P_2),(A_4,P_3)$. Hence $A_4$ is strictly better off. Besides, if $A_3$ also rejects $\{P_2,P_3\}$ as they are $A_3$'s two least favorite agents, the decentralized matching gives the outcome $(A_1,P_1),(A_2,P_2),(A_3,P_4),(A_4,P_3)$. Hence $A_3$ and $A_4$ are both strictly better off. 
Moreover, suppose there is a coordination mechanism among arms such that each arm only  accepts the most preferred agent. The decentralized matching gives the outcome $(A_1,P_4),(A_2,P_3),(A_3,P_1),(A_4,P_2)$, which is the arm-optimal stable matching.

\subsection{Comparison of multi-stage and single-stage matching}
\label{sec:multivssinglematching}

In this example, we show the gap between multi-stage welfare and single-stage welfare.
Suppose there are four arms $\mathcal A=\{A_1,A_2,A_3,A_4\}$ and  three agents $\mathcal P=\{P_1,P_2,P_3\}$. Agents have varied quotas: $q_1=2$ and $q_2=q_3=1$. Arms' attributes are given by $v_1=v_2=v_3=2, v_4=1$, and $e_{13}=e_{23}=e_{32}=0,e_{12}=e_{22}= e_{31}=0.5, e_{11}=e_{21}=e_{33}=1$, $e_{14}=0.2$, $e_{24}=0.5$, $e_{34}=0.8$.
The latent utilities and arms' true preferences are shown in Table \ref{table:eg5}.
Suppose each agent uses the straightforward strategy by pulling its most preferred arms up to the quota. Then the single-stage matching has the outcome
$(A_1,P_1),(A_2,P_1),(A_3,P_3)$. The multi-stage matching gives the outcome
$(A_1,P_1),(A_2,P_1),(A_3,P_3),(A_4,P_2)$. Hence $P_2$ is strictly better off in multi-stage matching as $P_2$'s welfare increases from $0$ to $1.5$ by changing from single-stage matching to multi-stage matching. 
On the other hand, $P_1$ and $P_3$ have the same welfare in single-stage and multi-stage matching. This result corroborates Proposition \ref{prof:benefitofmulstamat}.


\subsection{Supplementary results for real application}
\label{sec:chisquaredtest}
We give supplementary results to the real data analysis, where
the admission data is from the \emph{New York Times} ``The Choice" blog (available at
\href{https://thechoice.blogs.nytimes.com/category/admissions-data}{\color{blue}{https://thechoice.blogs.nytimes.com/category/admissions-data}}). 
Two colleges, Harvard and Yale, are excluded from the sample due to a significant proportion of missing values.

\subsubsection{Chi-squared test with FDR control}

\begin{figure}[ht]
    \centering
    \includegraphics[width=\textwidth]{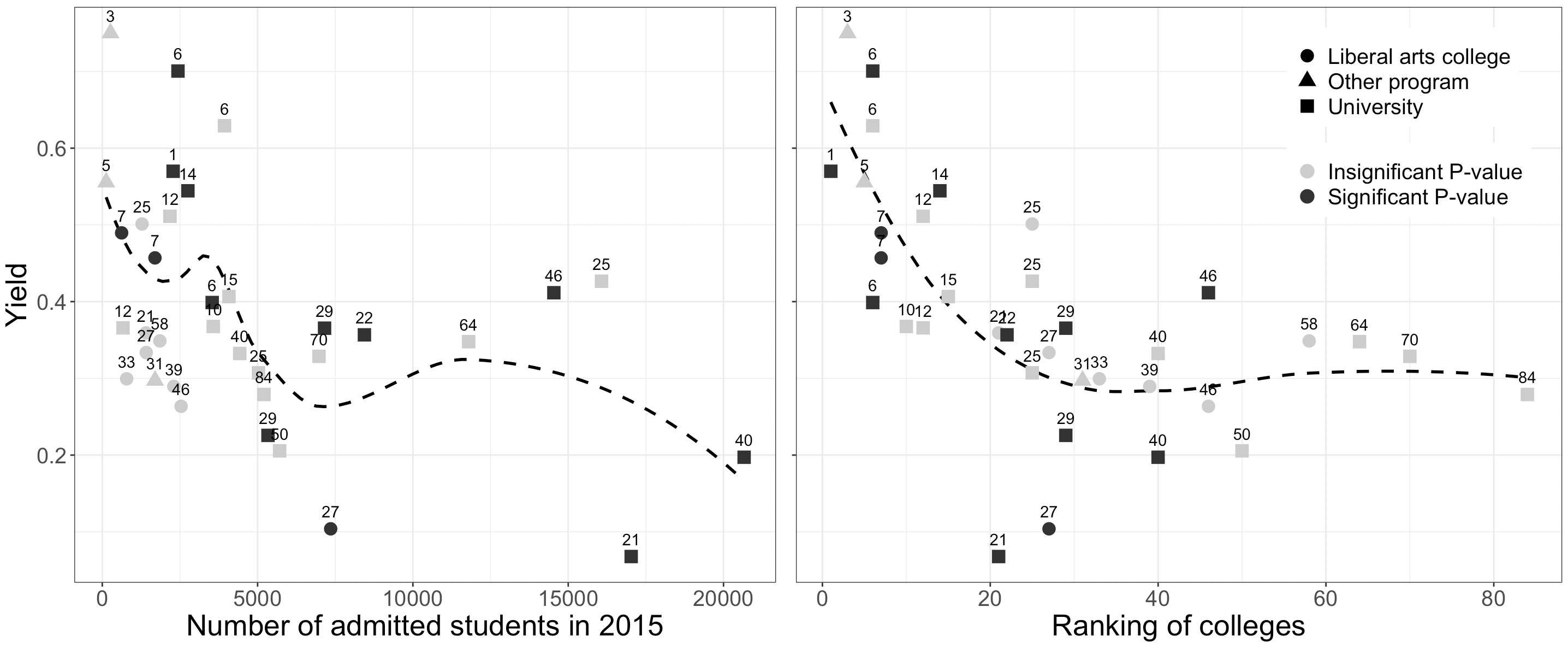}
    \caption{Regression of the yield on the size of admitted class and the ranking, respectively. 
    We fit the dashed curves using smoothing splines with the tuning parameter chosen by GCV.
     The labels $\{1,2,\ldots,35\}$ of each point indicates colleges' ranking according to \emph{U.S. News and World Report},
     where two (or more) colleges might tie in the ranking, and liberal arts colleges, national universities, and other undergraduate programs  are ranked separately within their categories.
     Gray and black points denote colleges with insignificant and significant $p$-values, respectively, in chi-squared tests under an FDR control. }
    \label{fig:nytdata}
\end{figure}
We test if the yields of colleges changed over  2015--17. The null hypothesis is that the state  is the same. We use a simultaneous chi-squared test for all colleges with the count data on accepted and enrolled students and under an FDR control  at a $.05$ significance level \cite{benjamini1995}. 
Figure \ref{fig:nytdata} shows that  colleges with large numbers of admitted students are likely to have significantly varied yields.  Moreover,  top-ranked national universities and liberal arts colleges are likely to have significantly varied yields.
This observation corroborates the uncertainty in applicants' preferences facing colleges.
Tables \ref{table:chisquare1} and \ref{table:chisquare2} report the $13$ colleges with significant $p$-values and  the $22$ colleges with insignificant $p$-values, respectively. 

\begin{table}[!ht]
\caption{$13$ chi-squared tests with significant $p$-value under the FDR control at the $.05$ significance level.
Colleges' ranking data are from \emph{U.S. News and World Report}. The "Y/N" means the use of the waiting list varied during 2015--17. }
\label{table:chisquare1}
\centering
\begin{tabular}{ l c c c c c c } 
\toprule
 & $p$-value & Category & Ranking & Waiting list \\[0.5ex] 
\midrule
Boston University & .0013 & National University & 40 & Yes \\  [0.3ex] 
Brown University  & .0012 & National University & 14 & No  \\ [0.3ex] 
Claremont McKenna College & .0003 & Liberal Arts College & 7 & Y/N \\ [0.3ex] 
College of Holy Cross & 2.20$E$-16 & Liberal Arts College & 27 & Yes   \\ [0.3ex]
Emory University & 2.20$E$-16 & National University & 21 & Yes \\  [0.3ex] 
Georgia Tech & .0022 & National University & 29 & Yes \\  [0.3ex] 
Middlebury College & .0065 & Liberal Arts College & 7 & Y/N \\  [0.3ex] 
Princeton University & 8.31$E$-12 & National University & 1 & Yes  \\  [0.3ex] 
Stanford University & 2.50$E$-06 & National University & 6 & Y/N  \\  [0.3ex] 
University of Chicago & 2.20$E$-06 & National University & 6 & Y/N \\  [0.3ex] 
University of Rochester & .0001 & National University & 29 & Y/N \\  [0.3ex] 
USC & 2.31$E$-11 & National University & 22 & No \\  [0.3ex] 
University of Wisconsin & .0008 & National University & 46 & Y/N  \\  [0.3ex] 
\bottomrule
\end{tabular}
\end{table}

\begin{table}[!ht]
\caption{$22$ chi-squared tests with insignificant $p$-value under the FDR control at the $.05$ significance level.
Colleges' ranking data are from \emph{U.S. News and World Report}. The "Y/N" means the use of the waiting list varied during 2015--17.}
\label{table:chisquare2}
\begin{center}
\begin{tabular}{ l c c c c c c} 
\toprule
 & $p$-value & Category & Ranking & Waiting list \\[0.5ex] 
\midrule
Babson College & .8994 & Other Program & 31 & Yes \\ [0.3ex] 
Barnard College  & .6159 & Liberal Arts College & 25 & Yes  \\ [0.3ex] 
Bates College  &  .0798 &  Liberal Arts College & 21 & Yes \\ [0.3ex] 
CalTech & .0584  &  National University &12 & Y/N \\ [0.3ex]
Carnegie Mellon University &.4988 &  National University & 25& Yes \\ [0.3ex]
College of William$\&$Mary &.2227 &  National University & 40 & Yes  \\ [0.3ex] 
Cooper Union &.9512  & Other Program & 3& Yes  \\ [0.3ex] 
Dartmouth College  &.2217 &  National University & 12 & Y/N \\  [0.3ex] 
Dickinson College  &.4727 & Liberal Arts College & 46& Y/N  \\  [0.3ex]  
Elon University & .6872 &  National University & 84& Y/N \\  [0.3ex] 
George Washington University  & .0309  &  National University & 70& Yes\\  [0.3ex] 
Johns Hopkins University & .1799&  National University & 10& Yes \\  [0.3ex] 
Kenyon College & .8012 & Liberal Arts College & 27 & Yes\\  [0.3ex] 
Lafayette College & .8719 & Liberal Arts College & 39  & Yes\\  [0.3ex] 
Olin College of Engineering  & .5317& Other Program & 5 & Y/N \\  [0.3ex] 
Rensselaer  Polytech & .0285&  National University & 50& Y/N \\  [0.3ex] 
Scripps College & .6511 & Liberal Arts College & 33 & Y/N \\  [0.3ex] 
St. Lawrence University  &.0587 & Liberal Arts College & 58& Yes \\  [0.3ex] 
University of Maryland  & .4438 &  National University & 64 & Y/N\\  [0.3ex] 
University of Michigan  & .0277&  National University & 25& Y/N \\ [0.3ex] 
University of Pennsylvania & .3665&  National University & 6 & Y/N\\  [0.3ex]       
Vanderbilt University &.7576  &  National University & 15& Y/N  \\ [0.3ex] 
\bottomrule
\end{tabular}
\end{center}
\end{table}

\subsubsection{Evidence on hierarchical structure}
\begin{figure}[ht]
    \centering
    \includegraphics[width=\textwidth]{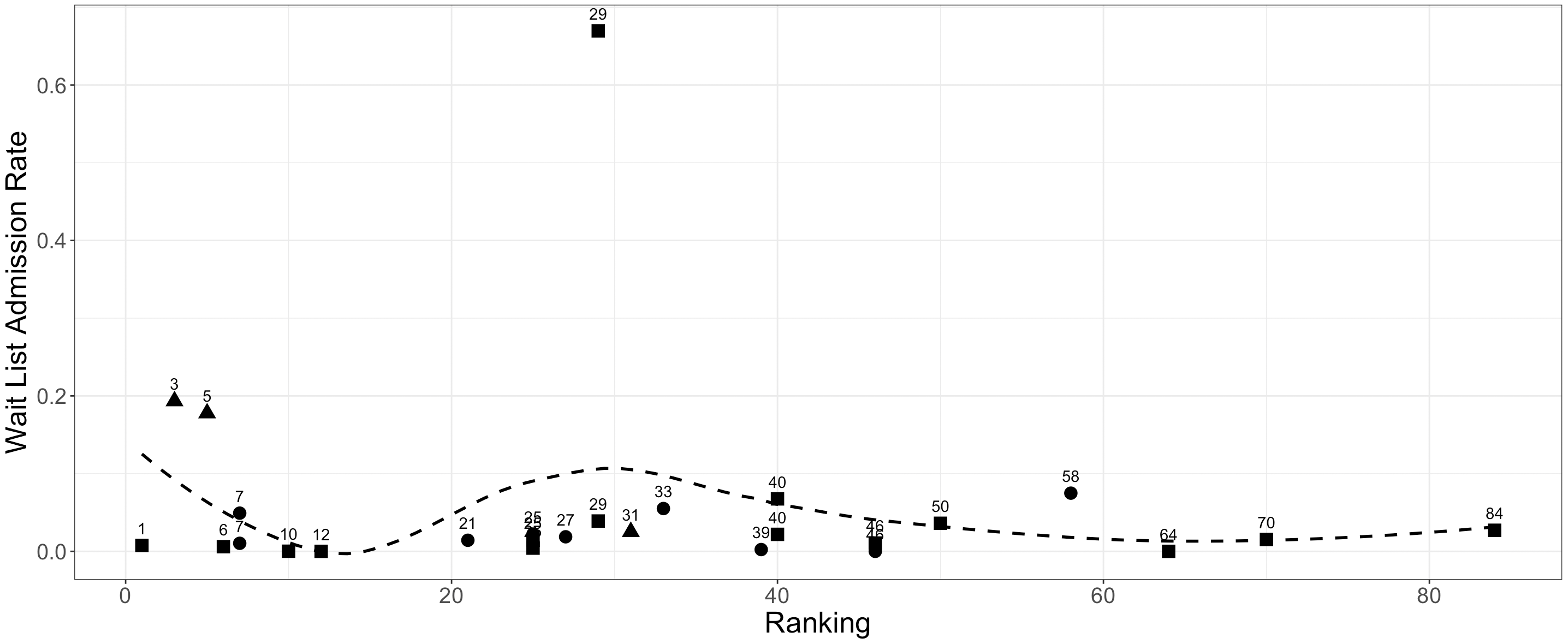} 
    \caption{The regression uses smoothing splines with the tuning by GCV.}
    \label{fig:NYTWaitAccept}
\end{figure}

We present the evidence on the hierarchical structure in the sense that students who were invited to the waiting list and remain available at a later stage are likely to be far worse than the admitted students at the regular admission stage. 
The report of National Association for College Admission Counseling \cite{nacac} shows that  the admission rate of the waiting list is significantly lower than that of regular admission. The top students in a college's waiting list, uncertain about their rankings in the list and whether the college would admit them later, may have accepted offers from their less preferred colleges.
We calculate the admission rate of the waiting list as follows:
\begin{equation*}
\frac{\text{the number of offers sent to wait-listed students}}{\text{ the total number of students invited to the waiting list}}.
\end{equation*}
Figure \ref{fig:NYTWaitAccept} reports that the majority  ($>77\%$) of  admission rate of the waiting list are below $5\%$, which result corroborates the existence of the hierarchical structure in college admissions with waiting lists. 




\section{Proofs}
\subsection{Proof of Theorem \ref{thm:reducvar}}
\label{sec:profofreducvar}

\subsubsection{Hierarchical structure}
We the exploit the underlying hierarchical structure of the optimization problem in Eq.~(\ref{eqn:optimalsolution}). 
For an arm set $\mathcal B_{i,k}\subseteq \{\mathcal A_k\setminus \cup_{l\leq k-1}\mathcal B_{i,l}\}$, its loss can be formulated by comparing its expected payoff to the expected payoff of $\bar{\mathcal B}_{i,k}$, where we suppose that $\cup_{k\in[K]}\bar{\mathcal B}_{i,k}$ achieves the optimal value $\bar{\mathcal U}_i$ in (\ref{eqn:optimalsolution}). Then the loss of $\mathcal B_{i,k}$ for any $k\in[K]$ becomes
\begin{equation}
\label{eqn:oeue}
\begin{aligned}
   \mathcal L_{i,k}[\mathcal B_{i,k}] & =   \mathbf 1\Big\{\sum_{j\in\mathcal B_{i,k}}\pi_{i,k}(s^*_{i,k},v_j)>q_i-\text{card}(\cup_{l\leq k-1}\mathcal C_{i,l})\Big\}\text{OE}[\mathcal B_{i,k}]\\
    &\quad \quad +\mathbf 1\Big\{\sum_{j\in\mathcal B_{i,k}}\pi_{i,k}(s_{i,k}^*,v_j)\leq q_{i}-\text{card}(\cup_{l\leq k-1}\mathcal C_{i,l})\Big\}\text{UE}[\mathcal B_{i,k}].
\end{aligned}
\end{equation}
Here the \emph{over-enrollment} (OE) loss in (\ref{eqn:oeue}) is defined as
\begin{equation*}
\begin{aligned}
\text{OE}[\mathcal B_{i,k}]  & \equiv \gamma_i\Big\{\sum_{j\in\mathcal B_{i,k}}\pi_{i,k}(s_{i,k}^*,v_j)+\text{card}(\cup_{l\leq k-1}\mathcal C_{i,l})-q_i\Big\} \\
&\quad\quad - \Big\{\sum_{j\in\mathcal B_{i,k}}(v_j+e_{ij})\pi_{i,k}(s_{i,k}^*,v_j) -\sum_{j\in\bar{\mathcal B}_{i,k}}(v_j+e_{ij})\pi_{i,k}(s_{i,k}^*,v_j)\Big\},\quad \forall k\in[K],
\end{aligned}
\end{equation*}
where we recall that penalty parameter $\gamma_i$ is defined in (\ref{eqn:totalutility}).
The \emph{under-enrollment} (UE) loss  in (\ref{eqn:oeue})  is given by
\begin{equation}
\label{eqn:ueb}
\begin{aligned}
& \text{UE}[\mathcal B_{i,k}]\\
&\equiv 
\begin{cases}
\rho_{i,k}[\sum_{j\in\bar{\mathcal B}_{i,k}}(v_j+e_{ij})\pi_{i,k}(s_{i,k}^*,v_j) -\sum_{j\in\mathcal B_{i,k}}(v_j+e_{ij})\pi_{i,k}(s_{i,k}^*,v_j)], &k\leq K-1,\\
\sum_{j\in\bar{\mathcal B}_{i,k}}(v_j+e_{ij})\pi_{i,K}(s_{i,k}^*,v_j) -\sum_{j\in\mathcal B_{i,k}}(v_j+e_{ij})\pi_{i,K}(s_{i,k}^*,v_j), & k=K,
\end{cases}
\end{aligned}
\end{equation}
where $\rho_{i,k}\in(0,1)$ is a discount factor for $k\leq K-1$. Note that $\rho_{i,k}<1$ is because $P_i$ can fill the remaining quota (if any) in subsequent stages of the matching process.
On the other hand, $\rho_{i,k}>0$ is due to the observation that the arms available at subsequent stages are likely to be worse than the arms available at the current stage. Specifically, we refer to this observation as the \emph{hierarchical structure} of the multi-stage matching and it is defined as follows: For any agent $P_i$, the $j$th best arm available at the subsequent stage has lower latent utility than the $j$th best arm available at the current stage, where $j\geq 1$. The hierarchical structure has been noted in college admissions with waiting lists  \citep{che2016}. 
Unlike the stages $k\leq K-1$, the last stage $k=K$ has the discount factor equals to $1$ since the agent cannot fill the remaining quota (if any) after the last stage.

The formulation in Eq.~(\ref{eqn:oeue}) allows one to study stage-wise optimal sets $\mathcal B_{i,k}$ that minimize the loss $\mathcal L_{i,k}$ for each $k\in[K]$. This makes the optimization problem easier compared to jointly finding $\mathcal B_{i,k}$ for all $k\in[K]$ such that $\cup_{k\in[K]}\mathcal B_{i,k}$ maximizes the expected payoff in (\ref{eqn:optimalsolution}). 

\subsubsection{Main proof of Theorem \ref{thm:reducvar}}
\begin{proof}
We introduce additional notations. 
Let $V_{i,k}(s_{i,k}^*, \mathcal B_{i,k})$ be the expected utility of arms from $\mathcal B_{i,k}\subseteq \{\mathcal A_k\setminus \cup_{l\leq k-1}\mathcal B_{i,l}\}$ for agent $P_i$ at stage $k\in[K]$. That is, 
\begin{equation*}
V_{i,k}(s_{i,k}^*,\mathcal B_{i,k}) \equiv \sum_{j\in\mathcal B_{i,k}}(v_j+e_{ij})\pi_{i,k}(s_{i,k}^*,v_j).
\end{equation*} 
Let $\mathcal N_{i,k}(s_{i,k}^*,\mathcal B_{i,k})$ be the expected number of arms in $\mathcal B_{i,k}$ accepting $P_i$. That is,
\begin{equation*}
\mathcal N_{i,k}(s_{i,k}^*,\mathcal B_{i,k}) \equiv \sum_{j\in\mathcal B_{i,k}}\pi_{i,k}(s_{i,k}^*,v_j).
\end{equation*}
By Lagrangian duality, the optimization of $\mathcal L_{i,k}[\mathcal B_{i,k}]$ in Eq.~(\ref{eqn:oeue}) can be reformulated to the constraint form:
\begin{equation*}
\begin{aligned}
& \underbrace{\max_{\mathcal B_{i,k}\subseteq \{\mathcal A_k\setminus \cup_{l\leq k-1}\mathcal B_{i,l}\}}\Big\{ V_{i,k}(s_{i,k}^*,\mathcal B_{i,k})-\gamma_i\max\{\mathcal N_{i,k}(s_{i,k}^*,\mathcal B_{i,k})+\text{card}(\cup_{l\leq k-1}\mathcal C_{i,l})-q_i,0\}\Big\}}_{\mathcal I_1},\\
& \quad\quad\quad\quad\quad\quad\quad\quad\quad\quad\quad\quad\quad\quad\quad\quad\quad\quad \text{s.t. }\underbrace{\text{UE}(\mathcal B_{i,k})\geq \eta_{i,k}'}_{\mathcal I_2},
\end{aligned}
\end{equation*}
Here  $\eta'_{i,k}>0$  is an appropriately chosen tolerance parameter for $k\leq K-1$, and $\eta_{i,K}'=0$.
The constraint $\mathcal I_2$ can be written as
\begin{equation}
\label{eqn:condi2}
     V_{i,k}(s_{i,k}^*,\mathcal B_{i,k})\leq V_{i,k}(s_{i,k}^*,\mathcal B_{i,k}^*) -\eta_{i,k}',\quad\forall s_{i,k}^*,
\end{equation}
where $ \mathcal B_{i,k}\subseteq\{\mathcal A_k\setminus\cup_{l\leq k-1}\mathcal B_{i,l}\}$.
Since  $\pi_{i,k}(\cdot,\cdot)$ is assumed to belong to an RKHS, $\pi_{i,k}(\cdot,\cdot)$ is bounded \citep{wahba1990spline}. By Hoeffding's bound, with probability at least $1-e^{-\epsilon},\forall \epsilon>0$,
\begin{equation*}
\begin{aligned}
  V_{i,k}(s_{i,k}^*,\mathcal B_{i,k}) 
   & < \E_{s_{i,k}^*}[V_{i,k}(s_{i,k}^*,\mathcal B_{i,k})] + \sqrt{2\epsilon\sum_{j\in\mathcal B_{i,k}}\delta^2_{i,k}(v_j)(v_j+e_{ij})^2}\\
    & < \E_{s_{i,k}^*}[V_{i,k}(s_{i,k}^*,\mathcal B_{i,k})] + \sqrt{2\epsilon}\sum_{j\in\mathcal B_{i,k}}\delta_{i,k}(v_j)(v_j+e_{ij}).
\end{aligned}
\end{equation*}
Hence a sufficient condition for Eq.~(\ref{eqn:condi2})  is to control 
\begin{equation}
\label{eqn:condi2'}
  \sum_{j\in\mathcal B_{i,k}}\delta_{i,k}(v_j)(v_j+e_{ij})<\eta_{i,k}'',\quad\text{for }\mathcal B_{i,k}\subseteq\{\mathcal A_k\setminus\cup_{l\leq k-1}\mathcal B_{i,l}\}.
\end{equation}
Here  $\eta_{i,k}''>0$ is a tolerance parameter  for $k\leq K-1$.
Both the $\mathcal I_1$ and Eq.~(\ref{eqn:condi2'}) are convex, and so by Lagrangian duality, they can be reformulated in the penalized form that finding $\mathcal B_{i,k}\subseteq \{\mathcal A_k\setminus \cup_{l\leq k-1}\mathcal B_{i,l}\}$ to maximize
\begin{equation*}
\begin{aligned}
&\sum_{j\in\mathcal B_{i,k}}(v_j+e_{ij})[\pi_{i,k}(s_{i,k},v_{j})-\eta_{i,k}\delta_{i,k}(v_j)] \\
&\quad\quad\quad\quad\quad -\gamma_i\max\{\mathcal N_{i,k}(s_{i,k}^*,\mathcal B_{i,k})+\text{card}(\cup_{l\leq k-1}\mathcal C_{i,l})-q_i,0\},
\end{aligned}
\end{equation*}
where $\eta_{i,k}>0$ for $k\leq K-1$ and $\eta_{i,K}=0$.  
This  completes the proof.
\end{proof}

\subsection{Proof of Theorem \ref{thm:expoptstrategy}} 
\label{subsec:optofdcb}

\subsubsection{Quantifying the cutoff for the greedy strategy}
Let $b_{i,k}$ be the value of $r$ of those arms on the cutoff. That is,  arms  on the cutoff satisfy  $b_{i,k} = (v+e_{i})[1-\eta_{i,k}\delta_{i,k}(v)\pi_{i,k}^{-1}(s_{i,k},v)] \geq 0$.
Let $\Pi_{i,k}(b_{i,k})$ be the expected number of arms in $\widehat{\mathcal B}_{i,k}(s_{i,k})$ that would accept $P_i$. That is,
\begin{equation*}
\begin{aligned}
&\Pi_{i,k}(b_{i,k}) \\
&= \sum_{j\in\mathcal A}\mathbf{1}\left(e_{ij}\geq \min\left\{\max\left\{b_{i,k}[1-\eta_{i,k}\delta_{i,k}(v_j)\pi_{i,k}^{-1}(s_{i,k},v_j)]^{-1}-v_j, 0\right\}, 1\right\}\right) \pi_{i,k}(s_{i,k},v_j).
\end{aligned}
\end{equation*}
If there exists some $b_{i,k}\geq 0$ such that $\Pi_{i,k}(b_{i,k}) = q_i - \text{card}(\cup_{l\leq k-1}\mathcal C_{i,l})$, we let $\widehat{b}_{i,k}(s_{i,k}) = b_{i,k}$ and the cutoff $\widehat{e}_{i,k}(s_{i,k}, v) = \min\{\max\{\widehat{b}_{i,k}(s_{i,k})[1-\eta_{i,k}\delta_{i,k}(v)\pi_{i,k}^{-1}(s_{i,k},v)]^{-1} - v,0\},1\}$. However, if there is no solution to $\Pi_{i,k}(b_{i,k}) = q_i - \text{card}(\cup_{l\leq k-1}\mathcal C_{i,l})$, we let
\begin{equation*}
\begin{aligned}
& b_{i,k}^{+}(s_{i,k}) = \underset{b_{i,k}\geq 0}{\arg\max}\left\{\Pi_{i,k}(b_{i,k})> q_i-\text{card}(\cup_{l\leq k-1}\mathcal C_{i,l})\right\}, \\
&b_{i,k}^{-}(s_{i,k}) = \underset{b_{i,k}\geq 0}{\arg\min}\left\{\Pi_{i,k}(b_{i,k})< q_i-\text{card}(\cup_{l\leq k-1}\mathcal C_{i,l})\right\}.
\end{aligned}
\end{equation*}
To choose between $b_{i,k}^+$ and $b_{i,k}^-$, it is necessary to balance the expected utility and the expected penalty for exceeding the quota due to pulling arms on the \emph{boundary}.
Define two cutoffs $e_{i,k}^{+}(s_{i,k}, v) \equiv \min\{\max\{b^{+}_{i,k}(s_{i,k})[1-\eta_{i,k}\delta_{i,k}(v)\pi_{i,k}^{-1}(s_{i,k},v)]^{-1}-v, 0\}, 1\}$ 
and $e_{i,k}^{-}(s_{i,k}, v) \equiv \min\{\max\{b^{-}_{i,k}(s_{i,k})[1-\eta_{i,k}\delta_{i,k}(v)\pi_{i,k}^{-1}(s_{i,k},v)]^{-1}-v, 0\}, 1\}$. The two cutoffs correspond to two sets, $\mathcal B^{+}_{i,k}(s_{i,k}) = \{j\ |\ e_{ij}\geq e_{i,k}^{+}(s_{i,k}, v_j)\}$ and $\mathcal B^{-}_{i,k}(s_{i,k}) = \{j \ | \ e_{ij}\geq e_{i,k}^{-}(s_{i,k}, v_j)\}$, respectively.
Consider the following condition for the arms on the boundary $\{\mathcal B^{+}_{i,k}(s_{i,k})\setminus \mathcal B^{-}_{i,k}(s_{i,k})\}$. This condition formalizes the comparison of the variational expected utility and the  expected penalty of exceeding the quota:
\begin{equation}
\label{eqn:jbim1+s}
\begin{aligned}
& \sum_{j\in \mathcal B^+_{i,k}(s_{i,k})\setminus\mathcal B^{-}_{i,k}(s_{i,k})}(v_j+e_{ij})[\pi_{i,k}(s_{i,k},v_j)-\eta_{i,k}\delta_{i,k}(v_j)]\\
&\quad\quad\quad\quad\quad\quad\geq  \gamma_i\sum_{j\in\mathcal B^+_{i,k}(s_{i,k})}\pi_{i,k}(s_{i,k},v_j)-\gamma_i[q_i-\text{card}(\cup_{l\leq k-1}\mathcal C_{i,l})].
\end{aligned}
\end{equation}
If (\ref{eqn:jbim1+s}) holds, 
let $\widehat{b}_{i,k}(s_{i,k})=b_{i,k}^{+}(s_{i,k})$ and otherwise, let $\widehat{b}_{i,k}(s_{i,k})=b_{i,k}^{-}(s_{i,k})$. Then the cutoff
\begin{equation}
\label{eqn:choiceofeimseq}
\widehat{e}_{i,k}(s_{i,k}, v) = \min\Big\{\max\Big\{\widehat{b}_{i,k}(s_{i,k})[1-\eta_{i,k}\delta_{i,k}(v)\pi_{i,k}^{-1}(s_{i,k},v)]^{-1}-v, 0\Big\}, 1\Big\}.
\end{equation}
Finally, using the greedy strategy, agent $P_i$ pulls arms from
\begin{equation*}
\begin{aligned}
\widehat{\mathcal B}_{i,k}(s_{i,k})& = \left.\left\{j \ \right|  \text{$A_j\in \{\mathcal A_k\setminus \cup_{l\leq k-1}\mathcal B_{i,l}\}$ with $(v_j,e_{ij})$ satisfying } e_{ij}\geq \widehat{e}_{i,k}(s_{i,k},v_j)\right\}\\
& = \left.\left\{j \ \right|  \text{$A_j\in \{\mathcal A_k\setminus \cup_{l\leq k-1}\mathcal B_{i,l}\}$ satisfying } r(A_j)\geq r_*\right\},
\end{aligned}
\end{equation*} 
where $r_*$ is the cutoff defined in Section \ref{sec:modifiedutility}.

\subsubsection{Main proof of Theorem \ref{thm:expoptstrategy}}
\begin{proof}
We define the function,
\begin{equation*}
\begin{aligned}
\text{UE}^\dagger  \equiv & \Big[\min_{j\in \mathcal B^-_{i,k}(s_{i,k})}(v_j+e_{ij})(1-\eta_{i,k}\pi_{i,k}^{-1}(s_{i,k},v_{j})\delta_{i,k}(v_{j}f))\Big]\\
& \quad\quad\quad \cdot \Big[q_i-\text{card}(\cup_{l\leq k-1}\mathcal C_{i,l})-\sum_{j\in \mathcal B^-_{i,k}(s_{i,k})}\pi_{i,k}(s_{i,k},v_j)\Big].
\end{aligned}
\end{equation*}
It is not hard to see that $\text{UE}^\dagger \geq 0$ and it equals $0$ if there is a continuum of arms and $\pi_{i,k}(\cdot,v)$ is continuous in $v$.
We divide the main proof of Theorem \ref{thm:expoptstrategy} into five steps.
\paragraph{Step 1.} We show that the optimal strategy prefers an arm with higher fit given the same score. Suppose that arms $A_{j_1}, A_{j_2}\in\{\mathcal A_k\setminus \cup_{l\leq k-1}\mathcal B_{i,l}\}$ have the same score $v_{j_1}=v_{j_2}$, but $A_{j_1}$ has a worse fit than $A_{j_2}$ to agent $P_i$. Now assume that $A_{j_1}$ was pulled by $P_i$ at stage $k$ but $A_{j_2}$ was not, that is, $A_{j_1}\in \widehat{\mathcal B}_{i,k}(s_{i,k}), A_{j_2}\not\in \widehat{\mathcal B}_{i,k}(s_{i,k})$.
Then the expected number of arms accepting $P_i$ is unchanged if $P_i$ replaces $A_{j_1}$ with $A_{j_2}$ in $\widehat{\mathcal B}_{i,k}(s_{i,k})$. On the other hand, 
since the loss function in Eq.~(\ref{eqn:utilitysequnm}) is strictly decreasing in fit $e_{ij}$, $P_i$ should pull $A_{j_2}$ instead $A_{j_1}$. This argument holds  regardless of strategies  of other agents. 

\paragraph{Step 2.} We show that the cutoff curve $\widehat{e}_{i,k}(s_{i,k},v)$  in Eq.~(\ref{eqn:choiceofeimseq}) is well-defined.  If the boundary $\{\mathcal B^{+}_{i,k}(s_{i,k})\setminus\mathcal B^{-}_{i,k}(s_{i,k})\}$ is not empty, then $P_i$ pulling an arm $A_j$ on the boundary yields the loss
\begin{equation*}
    \mathcal L_{i,k}^\dagger[A_j]\leq 0,
\end{equation*}
which justifies the condition specified by Eq.~(\ref{eqn:jbim1+s}).
Since $\widehat{e}_{i,k}(s_{i,k},v)\in[0,1]$, the cutoff curve is well-defined.

\paragraph{Step 3.} We show that the cutoff strategy of pulling arms from the set $\widehat{\mathcal B}_{i,k}(s_{i,k})$ is near-optimal. 
Let $\widetilde{\mathcal B}_{i,k}(s_{i,k})$ be any other arm set. Define the following mixed strategy:
\begin{equation*}
\sigma_{i,k}(s_{i,k},v, e_i;t) \equiv t\cdot\mathbf{1}\{(v,e_i) \in \widetilde{\mathcal B}_{i,k}(s_{i,k})\}+(1-t)\cdot\mathbf{1}\{(v,e_i) \in \widehat{\mathcal B}_{i,k}(s_{i,k})\},\quad\text{for } t\in[0,1].
\end{equation*}
The corresponding loss of the mixed strategy $\sigma_i$ is
\begin{equation*}
\begin{aligned}
\bar{\mathcal L}_{i,k}(t)  & =  \sum_{j\in\{\mathcal A_k\setminus \cup_{l\leq k-1}\mathcal B_{i,l}\}}(v_j+e_{ij})[\eta_{i,k}\delta_{i,k}(v_j)-\pi_{i,k}(s_{i,k},v_j)]\sigma_{i,k}(s_{i,k},v_j, e_{ij};t) \\
& + \gamma_i\max\Big\{\sum_{j\in\{\mathcal A_k\setminus \cup_{l\leq k-1}\mathcal B_{i,l}\}}\pi_{i,k}(s_{i,k},v_j)\sigma_{i,k}(s_{i,k},v_j,e_{ij};t)+\text{card}(\cup_{l\leq k-1}\mathcal C_{i,l})-q_i,0\Big\}.
\end{aligned}
\end{equation*}
It is clear that $\bar{\mathcal L}_{i,k}(t)$ is convex in $t$.
We discuss the local change $d \bar{\mathcal L}_{i,k}(0)/dt$ in three cases.

Case (I): Consider removing a single arm from $\widehat{\mathcal B}_{i,k}(s_{i,k})$. If the arm is from the non-empty boundary $\{\mathcal B^{+}_{i,k}(s_{i,k})\setminus\mathcal B^{-}_{i,k}(s_{i,k})\}$,  the condition specified by Eq.~(\ref{eqn:jbim1+s}) implies that the loss $\bar{\mathcal L}_{i,k}(t)$ increases if not pulling the arm. Moreover, by construction, any other arm $A_j$ in $\widehat{\mathcal B}_{i,k}(s_{i,k})$ satisfies
\begin{equation*}
\begin{aligned}
    & (v_j+e_{ij})[\pi_{i,k}(s_{i,k}^*,v_j)-\eta_{i,k}\delta_{i,k}(v_j)] > \widehat{b}_{i,k}(s_{i,k})\pi_{i,k}(s_{i,k}^*,v_j) \\
    &\geq \gamma_i\sum_{j'\in\widehat{\mathcal B}_{i,k}(s_{i,k})}\pi_{i,k}(s_{i,k},v_{j'})-\gamma_i[q_i-\text{card}(\cup_{l\leq k-1}\mathcal C_{i,l})].
\end{aligned}
\end{equation*}
Hence, removing $A_j$ from $\widehat{\mathcal B}_{i,k}(s_{i,k})$ results in a strict increase in $\bar{\mathcal L}_{i,k}(t)$.
We have $d \bar{\mathcal L}_{i,k}(0)/dt> 0$ in this case.
By the convexity of $\bar{\mathcal L}_{i,k}(t)$ in $t$,  we obtain
\begin{equation*}
\bar{\mathcal L}_{i,k}(1)= \bar{\mathcal L}_{i,k}(0)+\frac{d\bar{\mathcal L}_{i,k}(0)}{dt}(1-0)> \bar{\mathcal L}_{i,k}(0),
\end{equation*}

Case (II): Consider  adding a new arm with attributes $\{v_{j'},e_{ij'}\}$ to $\widehat{\mathcal B}_{i,k}(s_{i,k})$, where the new arm is not from the set $\mathcal B_{i,k}^{+}(s_{i,k})$. Denote by $\mathcal B_{i,k}'(s_{i,k})$ the new arm set with the added arm. Note that $P_i$ pulls a new arm only if the arm reduces the loss $\bar{\mathcal L}_{i,k}(t)$, that is,
\begin{equation}
\label{eqn:vj'eij'g}
\begin{aligned}
& (v_{j'}+e_{ij'})[\pi_{i,k}(s_{i,k},v_{j'})-\eta_{i,k}\delta_{i,k}(v_{j'})]\\
&\geq \gamma_i\sum_{j\in\mathcal B'_{i,k}(s_{i,k})}\pi_{i,k}(s_{i,k},v_j)-\gamma_i[q_i-\text{card}(\cup_{l\leq k-1}\mathcal C_{i,l})].
\end{aligned}
\end{equation}
Since the added new arm is not in $\mathcal B_{i,k}^{+}(s_{i,k})$ and $\sum_{j\in\mathcal B_{i,k}^{+}(s_{i,k})}\pi_{i,k}(s_{i,k},v_j)\geq q_i-\text{card}(\cup_{l\leq k-1}\mathcal C_{i,l})$,  we have
\begin{equation}
\label{eqn:sumjipisvjqii}
\begin{aligned}
& \sum_{j\in\mathcal B'_{i,k}(s_{i,k})}\pi_{i,k}(s_{i,k},v_j)-[q_i-\text{card}(\cup_{l\leq k-1}\mathcal C_{i,l})] \\
 &\geq \sum_{j\in\mathcal B'_{i,k}(s_{i,k})}\pi_{i,k}(s_{i,k},v_j)-\sum_{j\in\mathcal B_{i,k}^{+}(s_{i,k})}\pi_{i,k}(s_{i,k},v_j)\\
 &\geq \pi_{i,k}(s_{i,k},v_{j'})\\
 &\geq \pi_{i,k}(s_{i,k},v_{j'})-\eta_{i,k}\delta_{i,k}(v_{j'}).
\end{aligned}
\end{equation}
Because that $\gamma_i>\sup_{j\in\mathcal A}\{v_j+e_{ij}\}$ and $\eta_{i,k}\geq 0$,  the result in Eq.~(\ref{eqn:sumjipisvjqii}) is contradictory to Eq.~(\ref{eqn:vj'eij'g}). 
Hence, adding a new arm to $\widehat{\mathcal B}_{i,k}(s_{i,k})$ results in an increase in the loss  $\bar{\mathcal L}_{i,k}(t)$. Hence, $d \bar{\mathcal L}_{i,k}(0)/dt>0$ in this case.
By the convexity of $\bar{\mathcal L}_{i,k}(t)$ in $t$,  we obtain
\begin{equation*}
\bar{\mathcal L}_{i,k}(1)= \bar{\mathcal L}_{i,k}(0)+\frac{d\bar{\mathcal L}_{i,k}(0)}{dt}(1-0)> \bar{\mathcal L}_{i,k}(0),
\end{equation*}

Case (III): Consider  removing an arm with attributes $(v_j,e_{ij})$ from  $\widehat{\mathcal B}_{i,k}(s_{i,k})$ and simultaneously adding  new arms to  $\widehat{\mathcal B}_{i,k}(s_{i,k})$. Suppose that the new arms have attributes $(v_{j''},e_{ij''})$ and are from $\mathcal B_{i,k}''(s_{i,k})$.  If $\widehat{\mathcal B}_{i,k}(s_{i,k}) = \mathcal B_{i,k}^-(s_{i,k})$, then the new arms are not in $\mathcal B_{i,k}^{-}(s_{i,k})$ and by definition, 
\begin{equation*}
\begin{aligned}
& (v_{j''} + e_{ij''})[\pi_{i,k}(s_{i,k},v_{j''})-\eta_{i,k}\delta_{i,k}(v_{j''})]\pi^{-1}_{i,k}(s_{i,k},v_{j''})\\
& \quad\quad\quad\quad\quad\quad\leq \min_{j\in \mathcal B^-_{i,k}(s_{i,k})}\left\{(v_{j} + e_{ij})[\pi_{i,k}(s_i,v_j)-\eta_i\delta_{i,k}(v_j)]\pi^{-1}_{i,k}(s_{i,k},v_j)\right\}.
\end{aligned}
\end{equation*} 
Hence, 
\begin{equation}
\label{eqn:bdonb-}
\begin{aligned}
&\mathcal L_i^\dagger[\mathcal B^-_{i,k}(s_{i,k})] - \bar{\mathcal L}_{i,k}(1)  \\
&\leq 
 \sum_{j''\in\mathcal B_{i,k}''(s_{i,k})}(v_{j''}+e_{ij''})[\pi_{i,k}(s_{i,k},v_{j''})-\eta_{i,k}\delta_{i,k}(v_{j''})]\pi^{-1}_{i,k}(s_{i,k},v_{j''})\cdot \pi_{i,k}(s_{i,k},v_{j''})\\
&\leq \Big[\min_{j\in \mathcal B^-_{i,k}(s_{i,k})}(v_j+e_{ij})(1-\eta_{i,k}\pi_{i,k}^{-1}(s_{i,k},v_{j})\delta_{i,k}(v_{j}))\Big] \\
&\quad\quad\quad\quad\quad\quad\cdot\Big[q_i-\text{card}(\cup_{l\leq k-1}\mathcal C_{i,l})-\sum_{j\in \mathcal B^-_{i,k}(s_{i,k})}\pi_{i,k}(s_{i,k},v_j)\Big]\\
& = \text{UE}^\dagger.
\end{aligned}
\end{equation}
If $\widehat{\mathcal B}_{i,k}(s_{i,k}) = \mathcal B_{i,k}^+(s_{i,k})$, 
then by definition of $\mathcal B_{i,k}^+(s_{i,k})$
\begin{equation*}
\begin{aligned}
&\mathcal L_i^\dagger[\mathcal B^+_{i,k}(s_{i,k})]  - \bar{\mathcal L}_{i,k}(1) \leq \mathcal L_i^\dagger[\mathcal B^-_{i,k}(s_{i,k})] - \bar{\mathcal L}_{i,k}(1) \leq  \text{UE}^\dagger.
\end{aligned}
\end{equation*}
where the last inequality is by Eq.~(\ref{eqn:bdonb-}).
Hence,
\begin{equation*}
    \bar{\mathcal L}_{i,k}(0) - \bar{\mathcal L}_{i,k}(1) \leq  \text{UE}^\dagger.
\end{equation*}
Therefore, exchanging an arm in  $\widehat{\mathcal B}_{i,k}(s_{i,k})$  with arms not in  $\widehat{\mathcal B}_{i,k}(s_{i,k})$ could result in an increase in the loss $\bar{\mathcal L}_{i,k}(t)$ by at most  $\text{UE}^\dagger$. 
Combining the cases (I), (II), (III), we obtain that 
\begin{equation*}
    \mathcal L_i^\dagger[\widehat{\mathcal B}_{i,k}(s_{i,k})]\leq \min_{\mathcal B_{i,k}\subseteq \{\mathcal A_k\setminus \cup_{l\leq k-1}\mathcal B_{i,l}\}}\mathcal L_i^\dagger[\mathcal B_{i,k}] + \text{UE}^\dagger.
\end{equation*}

\paragraph{Step 4.} We prove the other direction of the inequality. Since $\widehat{\mathcal B}_{i,k}(s_{i,k})\subseteq \{\mathcal A_k\setminus \cup_{l\leq k-1}\mathcal B_{i,l}\}$,
\begin{equation*}
    \mathcal L_i^\dagger[\widehat{\mathcal B}_{i,k}(s_{i,k})]\geq \min_{\mathcal B_{i,k}\subseteq \{\mathcal A_k\setminus \cup_{l\leq k-1}\mathcal B_{i,l}\}}\mathcal L_i^\dagger[\mathcal B_{i,k}].
\end{equation*}
\paragraph{Step 5.} If there is a continuum of arms and $\pi_{i,k}(\cdot,v)$ is continuous in $v$, then there exists $b_{i,k}\geq 0$
such that
$\Pi_{i,k}(b_{i,k}) = q_i - \text{card}(\cup_{l\leq k-1}\mathcal C_{i,l})$, where $\Pi_{i,k}(b_{i,k})$ is defined in Section \ref{sec:near-optimal}:
\begin{equation*}
\Pi_{i,k}(b_{i,k}) = \sum_{j\in\mathcal A}\mathbf{1}\left(e_{ij}\geq \min\left\{\max\left\{b_{i,k}[1-\eta_{i,k}\delta_{i,k}(v_j)\pi_{i,k}^{-1}(s_{i,k},v_j)]^{-1}-v_j, 0\right\}, 1\right\}\right) \pi_{i,k}(s_{i,k},v_j).
\end{equation*}
Therefore, by definition, $\widehat{\mathcal B}_{i,k}(s_{i,k})=\mathcal B^+_{i,k}(s_{i,k}) = \mathcal B^-_{i,k}(s_{i,k})$,  and
\begin{equation*}
    q_i-\text{card}(\cup_{l\leq k-1}\mathcal C_{i,l})-\sum_{j\in \mathcal B^-_{i,k}(s_{i,k})}\pi_{i,k}(s_{i,k},v_j) = 0.
\end{equation*}
Hence $\text{UE}^\dagger = 0$. This completes the proof.
\end{proof}

\subsection{Proof of Theorem \ref{thm:optimalstatesseq}}
\label{sec:calminvarcdm}
\subsubsection{Main proof of Theorem \ref{thm:optimalstatesseq}}
\begin{proof}
We follow the proof arguments for Theorem 4 in \cite{dai2020learning}. The only difference is that here we 
define the following penalized expected utility and the expected number of arms:
\begin{equation*}
\begin{aligned}
V_{i,k}(s_{i,k}^*,\widehat{\mathcal B}_{i,k}) & \equiv \sum_{j\in\widehat{\mathcal B}_{i,k}}(v_j+e_{ij})[\pi_{i,k}(s_{i,k}^*,v_j)-\eta_{i,k}\delta_{i,k}(v_j)],\\
\mathcal N_{i,k}(s_{i,k}^*,\widehat{\mathcal B}_{i,k}) & \equiv \sum_{j\in\widehat{\mathcal B}_{i,k}}\pi_{i,k}(s_{i,k}^*,v_j).
\end{aligned}
\end{equation*} 
We omit the details for simplicity.
\end{proof}

\subsubsection{Calibration under the worst-case loss}
Besides the average-case loss in Theorem \ref{thm:optimalstatesseq}, we also consider the worst-case loss with respect to the unknown $s_{i,k}^*$.  Theorem \ref{prop:minimaxcalib} gives \emph{minimax calibration}, which calibrates $s_{i,k}$ to minimize the maximum loss $\max_{s_{i,k}^*}\{\mathcal L_{i,k}^\dagger[\widehat{\mathcal B}_{i,k}(s_{i,k})]\}$ over the unknown $s_{i,k}^*$.

\begin{theorem}
\label{prop:minimaxcalib}
The worse-case loss $\max_{s_i^*}\{\mathcal L_{i,k}^\dagger[\widehat{\mathcal B}_{i,k}(s_{i,k})]\}$ is minimized if $s_{i,k}\in[0,1]$ is chosen as the solution to
\begin{equation*}
\label{eqn:jbimmaxminseq}
\begin{aligned}
& \sum_{j\in\widehat{\mathcal B}_{i,k}(s_{i,k})}2(v_j+e_{ij})\delta_{i,k}(v_j)
+\sum_{j\in\widehat{\mathcal B}_{i,k}(0)}(v_j+e_{ij})\left[\pi_{i,k}(0,v_j)- \eta_{i,k}\delta_{i,k}(v_j)\right]\\
& \quad\quad = \sum_{j\in\widehat{\mathcal B}_{i,k}(1)}(v_j+e_{ij})\left[\pi_{i,k}(1,v_j)- \eta_{i,k}\delta_{i,k}(v_j)\right]  + \gamma_i\sum_{j\in\widehat{\mathcal B}_{i,k}(s_{i,k})}\pi_{i,k}(1,v_j)-\gamma_i q_i .
\end{aligned}
\end{equation*}
\end{theorem}
The Proof follows from Theorem 5 in \cite{dai2020learning}. 

\subsection{Proof of Proposition \ref{prop:fallthroughcracksstrategy}}
\label{sec:pffallthroughcracks}

\begin{proof}
Recall the cutoff parameter $\widehat{b}_{i,k}(s_{i,k})$ defined in Eq.~(\ref{eqn:choiceofeimseq}). Similarly, we define a cutoff parameter $b'_{i,k}(s_{i,k})$ for the linear cutoff:
$e_{i,k}'(s_{i,k},v) = \min\{\max\{b_{i,k}'(s_{i,k})-v,0\},1\}$ following three steps. First,
we define that
\begin{equation*}
    \Pi_{i,k}'(b_{i,k}) \equiv \sum_{j\in\mathcal A}\mathbf 1(e_{ij}\geq  \min\{\max\{b_{i,k}-v_j,0\},1\})\pi_{i,k}(s_{i,k},v_j).
\end{equation*}
If there exists $b_{i,k}\geq 0$ such that $\Pi_{i,k}'(b_{i,k})= q_i-\text{card}(\cup_{l\leq k-1}\mathcal C_{i,l})$, we let
$b'_{i,k}(s_{i,k})=b_{i,k}$. 
Second, if there is no solution to $\Pi_{i,k}'(b_{i,k})= q_i-\text{card}(\cup_{l\leq k-1}\mathcal C_{i,l})$, we let
\begin{equation*}
\begin{aligned}
& b_{i,k}^{+}(s_{i,k}) = \underset{b_{i,k}\geq 0}{\arg\max}\left\{\Pi'_{i,k}(b_{i,k})> q_i-\text{card}(\cup_{l\leq k-1}\mathcal C_{i,l})\right\}, \\
&b_{i,k}^{-}(s_{i,k}) = \underset{b_{i,k}\geq 0}{\arg\min}\left\{\Pi'_{i,k}(b_{i,k})< q_i-\text{card}(\cup_{l\leq k-1}\mathcal C_{i,l})\right\}.
\end{aligned}
\end{equation*}
Define that 
\begin{equation*}
    e_{i,k}^{+}(s_{i,k}, v) \equiv \min\{\max\{b^{+}_{i,k}(s_{i,k})-v, 0\}, 1\},
\end{equation*}
and 
\begin{equation*}
    e_{i,k}^{-}(s_{i,k}, v) \equiv \min\{\max\{b^{-}_{i,k}(s_{i,k})-v, 0\}, 1\}.
\end{equation*}
Then $e_{i,k}^{+}(s_{i,k}, v)$ and $e_{i,k}^{-}$
correspond to following sets respectively,
\begin{equation*}
    \mathcal B^{+}_{i,k}(s_{i,k}) = \{j\ |\ e_{ij}\geq e_{i,k}^{+}(s_{i,k}, v_j)\},
\end{equation*} and 
\begin{equation*}
    \mathcal B^{-}_{i,k}(s_{i,k}) = \{j \ | \ e_{ij}\geq e_{i,k}^{-}(s_{i,k}, v_j)\}.
\end{equation*} 
Consider the following condition for the arms on the boundary $\{\mathcal B^{+}_{i,k}(s_{i,k})\setminus \mathcal B^{-}_{i,k}(s_{i,k})\}$:
\begin{equation*}
\begin{aligned}
 \sum_{j\in \mathcal B^+_{i,k}(s_{i,k})\setminus\mathcal B^{-}_{i,k}(s_{i,k})}(v_j+e_{ij})\pi_{i,k}(s_{i,k},v_j)\geq  \gamma_i\sum_{j\in\mathcal B^+_{i,k}(s_{i,k})}\pi_{i,k}(s_{i,k},v_j)-\gamma_iq_i.
\end{aligned}
\end{equation*}
If the above condition holds, 
let $b'_{i,k}(s_{i,k})=b_{i,k}^{+}(s_{i,k})$ and otherwise, let $b'_{i,k}(s_{i,k})=b_{i,k}^{-}(s_{i,k})$. Third, we let the linear cutoff
\begin{equation*}
e'_{i,k}(s_{i,k}, v) = \min\{\max\{b'_{i,k}(s_{i,k})-v, 0\}, 1\}.
\end{equation*}
Now for any $\eta_{i,k}\geq 0$ and state $s_{i,k}$, the set of arms that have justified envy is 
\begin{equation*}
\mathcal V(\eta_{i,k},s_{i,k}) = \left\{(v,e_i) \  \left\vert \ \frac{\widehat{b}_{i,k}(s_{i,k})}{1-\eta_{i,k}\delta_{i,k}(v)\pi^{-1}_{i,k}(s_{i,k},v)} > v+e_{i}> b'_{i,k}(s_{i,k})  \right.\right\}. 
\end{equation*}
Hence the probability that an arm with attributes $(v,e_{})$ has justified envy is increasing in
\begin{equation}
\label{eqn:frcbik}
\frac{\widehat{b}_{i,k}(s_{i,k})}{1-\eta_{i,k}\delta_{i,k}(v)\pi_{i,k}^{-1}(s_{i,k},v)}.
\end{equation} 
Note that (\ref{eqn:frcbik}) is strictly increasing in the arm’s uncertainty level $\delta_{i,k}(v)\pi^{-1}_{i,k}(s_{i,k},v)$, 
the probability that an arm has justified envy is strictly increasing in the arm’s uncertainty level. 
\end{proof}


\subsection{Proof of Proposition \ref{thm:fairnessandeta}}

\begin{proof}
Adopting the proof in Section \ref{sec:pffallthroughcracks}, we note that
the number of arms having justified envy is
\begin{equation}
\label{eqn:armswithje}
\sum_{j\in\{\mathcal A_k^{T+1}\setminus\cup_{l\leq k-1}\mathcal B_{i,l}\}\cap \mathcal V(\eta_i,s)}\left[\frac{\widehat{b}_{i,k}(s_{i,k})}{1-\eta_{i,k}\delta_{i,k}(v_j)\pi_{i,k}^{-1}(s_{i,k},v_j)} - b'_{i,k}(s_{i,k})\right].
\end{equation}
The term in the bracket of Eq.~(\ref{eqn:armswithje}), i.e.,
\begin{equation*}
\frac{\widehat{b}_{i,k}(s_{i,k})}{1-\eta_{i,k}\delta_{i,k}(v_j)\pi_{i,k}^{-1}(s_{i,k},v_j)} - b'_{i,k}(s_{i,k})
\end{equation*} 
is strictly increasing in $\eta_{i,k}$.
Hence the number of arms having justified envy is strictly increasing in $\eta_{i,k}\geq 0$.  This completes the proof.
\end{proof}

\subsection{Proof of Proposition \ref{prof:benefitofmulstamat}}
\label{sec:benmulstage}

\begin{proof}
We show the improved welfare for agents by construction. 
Consider the strategy of an agent, for example, $P_i$ with $i \in[m]$.  Suppose that $P_i$ pulls arms at the first stage in multi-stage matching using the strategy that $P_i$ would have used in single-stage matching. 
All arms that would have accepted $P_i$ in single-stage matching accept $P_i$. The reason is that  arms have incomplete information on what other offers are coming in later stages. 
Hence, $P_i$ can achieve at least as well as its payoff from single-stage matching. Therefore, agents benefit from  multi-stage matching.
\end{proof}




\subsection{Proof of Proposition \ref{thm:explimprovs}}
\label{sec:pfofthmmincdm}

\begin{proof}

 \begin{figure}[ht]
    \centering
    \includegraphics[width=0.9\textwidth]{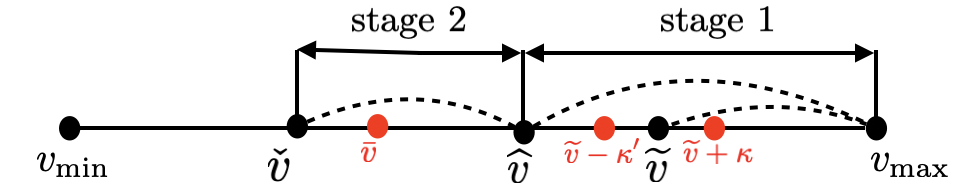}
    \caption{Cutoffs at the two stages.}
    \label{fig:threeagents}
\end{figure}

First, we consider the matching outcome of the straightforward strategy by pulling arms according to the latent utilities. 
Suppose that agents $P_1$ and $P_2$ use the CDM algorithm, which is a straightforward strategy and calibrates the uncertain state in the same way as LUB-CDM \citep{dai2020learning}.  The calibration in Theorem \ref{thm:optimalstatesseq} calibrates the  state parameters as $s=s_a$ for $P_1$ and $s=s_b$ for $P_2$. We note that worst-case calibration in Theorem \ref{prop:minimaxcalib} gives the same calibrations in this example.
Thus,  $P_1$ and $P_2$ pull the same set of arms at the first stage, where the arms' scores  $v\geq\widetilde{v}$  and the cutoff $\widetilde{v}$ satisfies
\begin{equation}
\label{eqn:numarmbyp1p2}
\sum_{j\in\mathcal A}\mathbf{1}(v_j\geq \widetilde{v})\cdot s_a\cdot (1-p^*)=q, \quad\text{and}\quad \sum_{j\in\mathcal A}\mathbf{1}(v_j\geq \widetilde{v})\cdot (1-s_b)\cdot (1-p^*)=q.
\end{equation}
Here the boundary arm set is assumed to be empty in Eq.~(\ref{eqn:numarmbyp1p2}).
Next, we consider $P_3$'s strategy. Arms with the scores worse than $\widetilde{v}$ will accept $P_3$ since if they accept $P_3$, they  get $u_3$ for sure, but if they reject $P_3$, they will at best  be pulled by $P_1$ or $P_2$ with probability $(1-p^*)$ and get the utility at most $u_1$, but $u_3>(1-p^*)u_1$. 
Suppose now $P_3$ pulls arms with the score $v\geq \widehat{v}$, where $\widehat{v}<\widetilde{v}$. By Eq.~(\ref{eqn:numarmbyp1p2}), there are total $(p^*)^2 q[s_a(1-p^*)]^{-1}$ of arms with $v\geq \widetilde{v}$ that are not pulled by $P_1$ or $P_2$ and they will accept $P_3$. Thus, we can quantify $\widehat{v}$ by letting it satisfy
\begin{equation*}
\sum_{j\in\mathcal A}\mathbf{1}( \widehat{v}\leq v_j<\widetilde{v}) = q\left[1-\frac{(p^*)^2}{s_a(1-p^*)}\right].
\end{equation*}
See an illustration of the cutoffs in Figure \ref{fig:threeagents}.
Then we analyze $P_1$'s expected payoff by using the CDM. If the true state is $s_b$, $P_1$ does not fill its capacity during the first stage and needs to pull more arms at the second stage. Suppose that $P_1$ pulls arms with $v\in[\check{v},\widehat{v})$ at the second stage, where $\check{v}$ satisfies 
\begin{equation}
\label{eqn:defcheckv}
\sum_{j\in\mathcal A}\mathbf{1}(\check{v}\leq v_j<\widehat{v})\cdot(1-p^*) = q - \sum_{j\in\mathcal A}\mathbf{1}(v_j\geq \widetilde{v})\cdot s_b\cdot(1-p^*).
\end{equation}
Hence, $P_1$'s expected payoff by using CDM is
\begin{equation}
\label{eqn:payoffofAnoexp}
\begin{aligned}
\mathcal U_1^{\text{CDM}}  = \frac{1}{2}(1-p^*)\left[\sum_{v_j\geq \widetilde{v}}v_j  + \sum_{\check{v}\leq v_j< \widehat{v}}v_j\right].
\end{aligned}
\end{equation}

We then consider the matching outcome of the LUB-CDM algorithm. Suppose that $P_1$ uses the LUB-CDM  while $P_2$ still uses the CDM. 
By Theorem \ref{thm:expoptstrategy}, $P_1$ pulls arms according to the ranking of the following quantity:
\begin{equation}
\label{eqn:varvjvarpisvj}
v_j\left[1 - \eta_{1,1}\cdot\frac{\delta_{1,1}(v_j)}{s}\right] = v_j\left[1 - \eta_{1,1}\cdot\frac{s_a-s_b}{2s}\right], 
\end{equation}
where $\delta_{1,1}(v) = \frac{1}{2}(s_a-s_b)$ in this example and
$\eta_{1,1}\geq 0$ is the regularization parameter defined in Theorem \ref{thm:reducvar}.
The calibration in  Theorem \ref{thm:optimalstatesseq} calibrates the state parameter as $s=s_a$ for $P_1$. Then $P_1$ pulls the arms with the score $v\in[\widetilde{v} - \kappa',\widetilde{v})\cup\{v\geq \widetilde{v}+\kappa\}$ and rejects those with $v\in[\widetilde{v},\widetilde{v}+\kappa)$.
Here the boundary arm set is assumed to be empty, and $\kappa,\kappa'$ satisfy
\begin{equation}
\label{eqn:sagtildevdelta}
\sum_{j\in\mathcal A}\mathbf{1}(\widetilde{v}-\kappa'\leq v_j< \widetilde{v})=\sum_{j\in\mathcal A}\mathbf{1}(\widetilde{v}\leq v_j< \widetilde{v}+\kappa)\cdot s_a.
\end{equation}
By Eq.~(\ref{eqn:varvjvarpisvj}), $\kappa$ and $\kappa'$ also need to satisfy that
\begin{equation}
\label{eqn:tildk'keta}
\widetilde{v} - \kappa' = (\widetilde{v}+\kappa)\left[1-\eta_{1,1}\cdot\frac{s_a-s_b}{2s_a}\right].
\end{equation}
Then we analyze $P_1$'s expected payoff by using the LUB-CDM. 
If the true state is $s_b$, $P_1$ needs to pull more arms at the second stage. 
Since the second stage is the last stage and by Theorem \ref{thm:reducvar}, it is optimal for $P_1$ to choose $\eta_{1,2}=0$, where the LUB-CDM coincides with the CDM. Suppose that $P_1$ pulls arms with $v\in[\bar{v},\widehat{v})$ at the second stage, where $\bar{v}$ satisfies
\begin{equation}
\label{eqn:defoverlinev}
\begin{aligned}
&\sum_{j\in\mathcal A}\mathbf{1}(\bar{v}\leq v_j<\widehat{v})\cdot(1-p^*) \\
&= q - \sum_{j\in\mathcal A}\mathbf{1}(v_j\geq \widetilde{v}+\kappa)\cdot s_b\cdot(1-p^*) - \sum_{j\in\mathcal A}\mathbf{1}(\widetilde{v}-\kappa'\leq v_j<\widetilde{v})\cdot(1-p^*) .
\end{aligned}
\end{equation}
Subtracting Eq.~(\ref{eqn:defoverlinev}) from Eq.~(\ref{eqn:defcheckv}), we obtain that
\begin{equation}
\label{eqn:sumjaacheckvjaa}
\begin{aligned}
\sum_{j\in\mathcal A}\mathbf{1}(\check{v}\leq v_j<\bar{v}) & =  \sum_{j\in\mathcal A}\mathbf{1}(\widetilde{v}-\kappa'\leq v_j<\widetilde{v}) - \sum_{j\in\mathcal A}\mathbf{1}(\widetilde{v}\leq v_j<  \widetilde{v}+\kappa)\cdot s_b \\
& = \sum_{j\in\mathcal A}\mathbf{1}(\widetilde{v}\leq v_j<  \widetilde{v}+\kappa)\cdot (s_a-s_b)>0.
\end{aligned}
\end{equation}
where the second equality is by Eq.~(\ref{eqn:sagtildevdelta}). Thus, $\bar{v}>\check{v}$, and
the $P_1$'s expected payoff by using the LUB-CDM is 
\begin{equation}
\label{eqn:payoffofAwithexp}
\begin{aligned}
\mathcal U^{\text{LUB-CDM}}_1=(1-p^*)\sum_{\widetilde{v}-\kappa'\leq v_j<\widetilde{v}}v_j + \frac{1}{2}(1-p^*)\left[\sum_{v_j\geq \widetilde{v}+\kappa}v_j + \sum_{\bar{v}\leq v_j<\widehat{v}}v_j \right].
\end{aligned}
\end{equation}

We now comparing the two expected payoffs in Eqs.~(\ref{eqn:payoffofAwithexp}) and (\ref{eqn:payoffofAnoexp}), respective. By taking the difference, we have
\begin{equation}
\label{eqn:u1minvardcmu1c}
\begin{aligned}
&\mathcal U^{\text{LUB-CDM}}_1 - \mathcal U_1^{\text{CDM}} \\
& = (1-p^*)\sum_{\widetilde{v}-\kappa'\leq v_j<\widetilde{v}}v_j   - \frac{1}{2}(1-p^*)\left[\sum_{\widetilde{v}\leq v_j<\widetilde{v}+\kappa}v_j  + \sum_{\check{v}\leq v_j< \bar{v}}v_j\right]\\
& > (1-p^*)(\widetilde{v}-\kappa')\sum_{j\in\mathcal A}\mathbf{1}(\widetilde{v}-\kappa'\leq v_j<\widetilde{v})\\
&\quad
- (\widetilde{v}+\kappa)\sum_{j\in\mathcal A}\mathbf{1}(\widetilde{v}\leq v<\widetilde{v}+\kappa) 
-\bar{v}\sum_{j\in\mathcal A}\mathbf{1}(\check{v}\leq v<\bar{v})\\
& = [(\widetilde{v}-\bar{v})(s_a-s_b)-(2\kappa's_a+\kappa)]\sum_{j\in\mathcal A}\mathbf{1}(\widetilde{v}\leq v_j<\widetilde{v}+\kappa)\\
& = \left\{(s_a-s_b)\left[(1-\eta_{1,1})\widetilde{v}-\bar{v}\right] + [2s_a-\eta_{1,1}(s_a-s_b)-1]\kappa\right\}\sum_{j\in\mathcal A}\mathbf{1}(\widetilde{v}\leq v_j<\widetilde{v}+\kappa),
\end{aligned}
\end{equation}
where the second equality is due to Eqs.~(\ref{eqn:sagtildevdelta}) and (\ref{eqn:sumjaacheckvjaa}), and the last equality is by Eq.~(\ref{eqn:tildk'keta}). For sufficiently small $\kappa$ and $\eta_{1,1}$, we have 
\begin{equation*}
    \mathcal U^{\text{LUB-CDM}}_1 > \mathcal U_1^{\text{CDM}}.
\end{equation*}

Last, we quantify the improvement of the expected payoff. 
From Eq.~(\ref{eqn:sagtildevdelta}), $\kappa$ satisfies that
\begin{equation*}
\sum_{j\in\mathcal A}\mathbf{1}\left((\widetilde{v} + \kappa)\left(1-\eta_{1,1}\frac{s_a-s_b}{2s_a}\right)\leq v_j< \widetilde{v}\right)=\sum_{j\in\mathcal A}\mathbf{1}(\widetilde{v}\leq v_j< \widetilde{v}+\kappa)\cdot s_a.
\end{equation*}
Suppose that $v_j$ is uniformly distributed, we have a first-order approximation of the above equations:
\begin{equation*}
\widetilde{v} - (\widetilde{v} + \kappa)\left(1-\eta_{1,1}\frac{s_a-s_b}{2s_a}\right) = \kappa s_a,
\end{equation*}
which implies that 
\begin{equation*}
 \kappa = \frac{\widetilde{v}(s_a-s_b)\eta_{1,1}}{2s_a(1+s_a)-(s_a-s_b)\eta_{1,1}} = O(\eta_{1,1})
\end{equation*}
Plugging it  to Eq.~(\ref{eqn:u1minvardcmu1c}) suggests that a sufficient condition for $\mathcal U^{\text{LUB-CDM}}_1 > \mathcal U_1^{\text{CDM}}$ is
\begin{equation}
\label{eqn:eta2sastilbar2sa}
\eta_{1,1} < \frac{2s_a}{s_a-s_b}\cdot\frac{(1+s_a)(s_a-s_b)(\widetilde{v}-\bar{v})}{(s_a-s_b)(\widetilde{v}-\bar{v})+ (2s_a^2+1)\widetilde{v}}.
\end{equation}
By the condition that $\kappa'>0$, we have
\begin{equation}
\label{eqn:eta12satildev}
\eta_{1,1}<\frac{2s_a}{s_a-s_b}(1+s_a-\widetilde{v}).
\end{equation}
Under Eqs.~(\ref{eqn:eta2sastilbar2sa}) and (\ref{eqn:eta12satildev}), and noting that,
\begin{equation*}
\sum_{j\in\mathcal A}\mathbf{1}(\widetilde{v}\leq v_j<\widetilde{v}+\kappa) = O(\kappa) = O(\eta_{1,1}),
\end{equation*}
we have  that,
\begin{equation*}
\mathcal U^{\text{LUB-CDM}}_1 - \mathcal U_1^{\text{CDM}} = O(\eta_{1,1}).
\end{equation*}
This completes the proof.
\end{proof}

\end{document}